\documentclass[aps,prd, notitlepage, onecolumn,superscriptaddress,floatfix,letterpaper,nofootinbib, longbibliography]{revtex4-1}
\usepackage{amsmath, amssymb, graphics, setspace}
\usepackage{graphicx}
\usepackage{mathtools}
\usepackage{amsthm}

\usepackage[usenames, dvipsnames]{color}
\usepackage[svgnames]{xcolor}
\usepackage[colorlinks,citecolor=RoyalBlue, urlcolor=RoyalBlue, linkcolor=RoyalBlue ]{hyperref} 

\usepackage{sseq}
\usepackage[all,cmtip]{xy}

\usepackage{tikz}
\usetikzlibrary{fit,matrix}
\usetikzlibrary{decorations.markings}
\usetikzlibrary{tikzmark,decorations.pathreplacing,positioning}

\newcommand{\mathsym}[1]{{}}
\newcommand{\unicode}[1]{{}}

\newcommand{\Sec}[1]{Sec.~\ref{#1}}

\newcommand{\Table}[1]{Table \ref{#1}} 

\newcommand{\bea}{\begin{eqnarray}}
\newcommand{\eea}{\end{eqnarray}}

\newcommand{\ii}{\hspace{1pt}\mathrm{i}\hspace{1pt}}
\newcommand{\dd}{\hspace{1pt}\mathrm{d}}

\newcommand{\Spin}{{\rm Spin}}

\newcommand{\U}{{\rm U}}

\newcommand{\TP}{{\rm TP}}

\newcommand{\cD}{\mathcal{D}}

\newcommand{\Z}{\mathbb{Z}}

\newcommand{\R}{\mathbb{R}}

\newcommand{\Tr}{\mathrm{Tr}}

\newcommand{\rF}{\rm{F}}
\newcommand{\diag}{\rm{diag}}
\newcommand{\Kthree}{\mathrm{K3}}
\newcommand{\Enriques}{\mathrm{E}}

\DeclareMathOperator{\Ext}{Ext}
\DeclareMathOperator{\rank}{rank}
\DeclareMathOperator{\sign}{sign}

\newtheorem{theorem}{Theorem}

\newtheorem{remark}{Remark}

\newcommand{\Hom}{\rm{Hom}}

\begin{document}

\title{Anomaly of 4d Weyl Fermions with Discrete Symmetries}
\author{Zheyan Wan}
\email{wanzheyan@bimsa.cn}
\affiliation{Beijing Institute of Mathematical Sciences and Applications, Beijing 101408, China}

\begin{abstract}
We derive explicit anomaly-index formulas for four-dimensional Weyl fermions
charged under the finite symmetries
$\mathrm{Spin}\times\mathbb Z_n$ and
$\mathrm{Spin}\times_{\mathbb Z_2^{\mathrm F}}\mathbb Z_{2m}^{\mathrm F}$.
The strategy is to start from the standard perturbative anomaly indices for
$\mathrm{Spin}\times\mathrm U(1)$ and
$\mathrm{Spin}\times_{\mathbb Z_2^{\mathrm F}}\mathrm U(1)=\mathrm{Spin}^c$,
and then restrict the continuous $\mathrm U(1)$ symmetry to a finite cyclic
subgroup.  On the level of invertible field theories this gives natural
homomorphisms
$$
\mathrm{TP}_5(\mathrm{Spin}\times\mathrm U(1))
\longrightarrow
\mathrm{TP}_5(\mathrm{Spin}\times\mathbb Z_n),\quad
\mathrm{TP}_5(\mathrm{Spin}^c)
\longrightarrow
\mathrm{TP}_5(\mathrm{Spin}\times_{\mathbb Z_2^{\mathrm F}}\mathbb Z_{2m}^{\mathrm F}).
$$
We compute these maps explicitly by evaluating reduced $\eta$-invariants on
geometric representatives of the finite anomaly groups.  For
$\mathrm{Spin}\times\mathbb Z_n$, the relevant backgrounds are the
five-dimensional lens-space bundle $X(n;1,1)$ and the product
$L(n;1)\times\mathrm{K3}$.  For
$\mathrm{Spin}\times_{\mathbb Z_2^{\mathrm F}}\mathbb Z_{2m}^{\mathrm F}$, the
relevant backgrounds are $L(m;1,1,1)$ and, depending on the parity of $m$,
either $L(m;1)\times\mathrm{Enriques}$ or $L(m;1)\times\mathrm{K3}$.

The output is a pair of integer-valued anomaly indices for each finite
symmetry.  These indices are normalized in the cyclic factors of the finite
anomaly group, so they can be used directly in anomaly-cancellation checks for
fermions with discrete gauge or global symmetries.  They also provide the input
for symmetry-extension applications to topologically ordered quantum dark
matter: in Standard-Model examples, the indices identify the discrete
${\bf B}+{\bf L}$ mixed gauge--gravitational anomaly left by missing sterile
right-handed neutrinos and the finite $K$-gauge topological sector that can
match it.
\end{abstract}

\maketitle

\tableofcontents

\bigskip

\section{Introduction}

\subsection{Motivation}

Anomalies constrain which symmetries can be consistently imposed in a quantum
field theory.  For four-dimensional Weyl fermions with continuous gauge
symmetry, the perturbative anomaly is described by the familiar six-dimensional
index polynomial.  Discrete symmetries require a slightly different treatment.
Although a discrete gauge field has no ordinary de Rham field strength, a
fermion coupled to a finite background can still have a global anomaly.  Such an
anomaly is naturally measured by a five-dimensional invertible field theory, or
equivalently by a bordism invariant.

This paper derives explicit anomaly indices for four-dimensional Weyl fermions
with the finite symmetries
\bea
\Spin\times\Z_n,
\qquad
\Spin\times_{\Z_2^{\rF}}\Z_{2m}.
\eea
The first case describes an ordinary spin theory with an independent cyclic
charge symmetry.  The second case is the finite analogue of a $\Spin^c$
symmetry: the order-two element in $\Z_{2m}$ is identified with fermion
parity.  Our main goal is to express the resulting global anomalies in concrete
integer coordinates of the finite anomaly groups, so that the formulas can be
used directly in anomaly-cancellation tests.

One concrete application, developed in Refs.~\cite{2502.21319,2512.25038,2605.26202}, is
to use these finite anomaly coordinates as the input for symmetry extension. If
a set of Weyl fermions has a nonzero 't Hooft anomaly for a finite symmetry
$G$, but this anomaly pulls back to zero under an extension
$1\to K\to G_{\rm Tot}\to G\to1$, then a $K$-gauge topological field theory can
carry the same $G$ anomaly in the infrared.  In the topologically ordered dark
matter application of Ref.~\cite{2512.25038}, the general symmetry-extension
statements are organized prime by prime. For $G=\Spin\times_{\Z_2^{\rF}}\Z_{2m}^{\rF}$ and $m=2^p\cdot 3^r\cdot s$ with $2\nmid s$ and $3\nmid s$, the
$2$-primary anomaly is canceled for $2^p$ (or $2^{p+1}$) charge-one Weyl
fermions, using a minimal $K=\Z_4$ (or $K=\Z_2^{\rF}$) extension. The
$3$-primary anomaly can be canceled for $3^r$ charge-one Weyl fermions by a minimal
$K=\Z_{3}$ extension, while an $s$-primary component is canceled
by a trivial $K=\Z_1$ extension for $s$ charge-one Weyl fermions.  In the Standard Model application,
$G$ is a discrete baryon-plus-or-minus-lepton symmetry combined with spacetime
spin structure, and the fermions whose anomaly is to be matched are the sterile
right-handed neutrinos absent from the low-energy spectrum.

The later work~\cite{2605.26202} applies this idea to a generalized Standard
Model with $N_c$ colors and $N_f$ families, but the anomaly-trivializing kernel
is first a cyclic group $K=\Z_N$ determined by the finite anomaly and by $N_f$,
not by the color number $N_c$.  For the discrete $({\bf B}+{\bf L})$ anomaly one
finds minimal extension orders $N=1,3,4,12$ according to the $2$- and
$3$-primary parts of $N_f$.  Only after imposing the additional color-center
matching condition $N=N_c$ can this $\Z_N$ gauge group be identified with the
center of $\mathrm{SU}(N_c)$.  In the odd color-matched branch, the odd-color, fermionic-baryon condition and minimality then select
$N=N_c=N_f=3$.  Thus the formulas derived below are not only a classification of
finite-symmetry anomalies; they are the arithmetic data needed to test whether
missing chiral fermions can be replaced by a symmetry-preserving topologically
ordered sector, and when that sector can also be interpreted as a color-center
TQFT.

\subsection{Previous Work}

Discrete gauge anomalies have been studied from several complementary
viewpoints.  Early work by Ibáñez and Ross \cite{Ibanez1991}, Preskill et al.
\cite{Preskill1991}, and Banks and Dine \cite{Banks1992} emphasized anomaly
cancellation conditions for finite symmetries and their relation to underlying
continuous gauge symmetries.  Ibáñez's later analysis \cite{Ibanez1993} gave a
more detailed account of the discrete anomaly constraints used in particle
physics model building.

More recently, Hsieh \cite{1808.02881} revisited discrete gauge anomalies using
bordism and $\eta$-invariants, while García-Etxebarria and Montero
\cite{1808.00009} developed the Dai--Freed viewpoint on global anomalies in
particle physics.  The mathematical tools underlying these computations,
including $\eta$-invariants of lens spaces and related fixed-point formulas,
go back to work of Gilkey and collaborators
\cite{Gilkey1984,Gilkey1987,Gilkey1996,Gilkey1997,Gilkey1999,Gilkey2018}.
The general classification of invertible field theories by bordism is due to
Freed and Hopkins \cite{1604.06527}.

There are also important connections with anomaly matching, non-Abelian
discrete symmetries, and fermionic symmetry-protected topological phases.
Examples include discrete anomaly matching \cite{Csaki1998}, non-Abelian
discrete anomalies \cite{Araki2008}, and finite-group or fermionic SPT phases
\cite{1405.7689,1406.7329,1607.01873,1812.11959,1909.08775,Cordova-Ohmori,Cheng:2024awi-2411.05786}.
The anomaly constraints for theories with
$\Spin\times\Z_n$ and
$\Spin\times_{\Z_2^{\rF}}\Z_{2m}$ symmetry were further studied in
\cite{Cordova-Ohmori,Cheng:2024awi-2411.05786}.  Motivated by applications to
topological quantum dark matter and the Standard Model family puzzle, related
uses of these anomaly constraints appear in
\cite{2302.14862,2501.00607,2502.21319,2512.25038,2605.26202}.

\subsection{Contribution of This Paper}

The contribution of this paper is to give a direct bridge from the continuous
$\U(1)$ anomaly polynomial to finite-symmetry global anomaly indices.  The
bridge is implemented by the restriction homomorphisms
\bea
\TP_5(\Spin\times\U(1))\longrightarrow \TP_5(\Spin\times\Z_n),
\qquad
\TP_5(\Spin^c)\longrightarrow
\TP_5(\Spin\times_{\Z_2^{\rF}}\Z_{2m}).
\eea
We compute these maps explicitly by evaluating $\eta$-invariants on geometric
generators of the finite anomaly groups.

For $\Spin\times\Z_n$, the relevant geometric representatives are the
lens-space bundle $X(n;1,1)$, defined in \eqref{eq:lens-space-bundle}, and
the product $L(n;1)\times\Kthree$, where $\Kthree$ is the K3 surface.  The
first representative detects the pure discrete gauge anomaly, and the second
detects the mixed gauge--gravitational anomaly after a change of basis.  This
recovers and refines the formulas of \cite{1808.02881}, using the fixed-point
formulas of \cite{Gilkey1984,Gilkey1987,Gilkey1997}.

For $\Spin\times_{\Z_2^{\rF}}\Z_{2m}$, the relevant representatives are
$L(m;1,1,1)$ and either $L(m;1)\times\Enriques$ for even $m$, where
$\Enriques$ denotes the Enriques surface, or $L(m;1)\times\Kthree$ for odd
$m$.  This gives the finite $\Spin^c$-type anomaly indices and clarifies the
normalization of the second cyclic summand relative to
\cite{1808.02881,1808.00009,Gilkey1996,Gilkey2018}.

Here and below, $\TP_r(G)$ denotes the group of deformation classes of
$r$-dimensional invertible topological field theories with symmetry $G$, in
the sense of Freed--Hopkins \cite{1604.06527}.  A useful way to interpret this
notation is the following: an anomaly of a $d$-dimensional quantum field theory
with symmetry $G$ is represented by a $(d+1)$-dimensional invertible field
theory with the same symmetry.  Thus deformation classes of anomalies of
$d$-dimensional theories with symmetry $G$ are classified by $\TP_{d+1}(G)$.
Since this paper studies four-dimensional Weyl fermions, the relevant anomaly
groups are five-dimensional groups $\TP_5(G)$.

The group $\TP_r(G)$ is controlled by the $G$-bordism groups $\Omega_r^G$ and
$\Omega_{r+1}^G$.  In this paper $\Omega_r^G$ means
\bea
\Omega_r^G
=
\{\text{closed }r\text{-manifolds }M\text{ equipped with a }G\text{-structure}\}/\text{bordism}.
\eea
Freed and Hopkins show that there is a split short exact sequence
\bea
0\to\Ext^1(\Omega_r^G,\Z)\to\TP_r(G)\to\Hom(\Omega_{r+1}^G,\Z)\to0.
\eea
Thus the torsion information in $\TP_r(G)$ is detected by
$\Omega_r^G$, while the free part is detected by $\Omega_{r+1}^G$.  In the
five-dimensional finite-symmetry cases considered below, the relevant groups
are finite, and we identify the torsion anomaly group $\TP_5(G)$ with the
corresponding finite bordism group $\Omega_5^G$ after choosing the geometric
generators used in Sections \ref{sec:Spin-U1} and \ref{sec:Spinc}.

The final formulas for the two finite symmetries are given in
\eqref{eq:Spin-Zn-rewrite} and \eqref{eq:Spin-Z2m-rewrite}.  A compact summary
appears in \Table{table:summary}.

\subsection{Main results and conventions}

For ease of reference, we state the two final anomaly-index formulas at the beginning. The derivations, including the construction of the manifold generators and the evaluation of the relevant $\eta$-invariants, are given in the later sections. Throughout the paper, all $\eta$-invariants are understood modulo $\Z$ when they appear in the phase $\exp(-2\pi\ii\eta)$. We write the anomaly groups additively when discussing bordism generators and multiplicatively when discussing the associated invertible field theories.

We also fix the notation for bordism classes before it appears in the theorem
statements.  If $M$ is a closed manifold equipped with the relevant tangential
structure and background gauge field, then
$[M]$ denotes its bordism class.  A minus sign in front of a manifold denotes
orientation reversal: $-M$ is the same manifold with the reversed orientation
and the induced background structure.  Equivalently, in the bordism group,
$-[M]=[-M]$.  A plus sign between manifolds or between their bordism classes
means disjoint union, which is the group operation in bordism.  Thus, for
example,
\bea
[M]+[N]=[M\sqcup N],
\eea
and $k[M]$ means the disjoint union of $k$ copies of $M$, with the coefficient
understood modulo the order of the relevant cyclic summand when the bordism
group is finite.  These conventions apply to the formulas such as
$[M^5]=k_1[M_1^5]+k_2[M_2^5]$ below.

We now explain the geometric symbols $M_i^5$ which occur in the theorem
statements.  For the symmetry $\Spin\times\Z_n$, the two chosen
five-dimensional generators are
\bea
[M_1^5]&:=&[X(n;1,1)],\cr
[M_2^5]&:=&-[L(n;1)\times\Kthree]-6[X(n;1,1)].
\eea
Here $X(n;1,1)$ is the lens-space bundle defined in
\eqref{eq:lens-space-bundle}, while $L(n;1)\times\Kthree$ is the product of
the one-dimensional lens space with the K3 surface.  The second formula is an
identity in the bordism group: it means the bordism class represented by the
orientation reversal of $L(n;1)\times\Kthree$, together with six oppositely
oriented copies of $X(n;1,1)$.  The corresponding integer-valued anomaly
indices in Theorem \ref{thm:main-spin-zn} are obtained from the
$\R/\Z$-valued $\eta$-invariants of these generators by multiplying by the
orders of the two cyclic factors:
\bea
\nu_1'(n,q)&=&a_n\eta(M_1^5,\rho_q),\cr
\nu_2'(n,q)&=&b_n\eta(M_2^5,\rho_q).
\eea
Thus $M_1^5$ is the background used to measure the first cyclic coordinate,
while $M_2^5$ is the background used to measure the second cyclic coordinate.
Equivalently, if a general background has class
$[M^5]=k_1[M_1^5]+k_2[M_2^5]$, then its anomalous phase is obtained by the
additivity of the $\eta$-invariant from the two numbers $\nu_1'(n,q)$ and
$\nu_2'(n,q)$ as in \eqref{eq:Spin-Zn-rewrite}.

For the finite $\Spin^c$-type symmetry
$\Spin\times_{\Z_2^{\rF}}\Z_{2m}$, the two chosen five-dimensional generators
are
\bea
[M_3^5]&:=&[L(m;1,1,1)],\cr
[M_4^5]&:=&-[\Enriques\times L(m;1)]+12[L(m;1,1,1)],\qquad m\text{ even},\cr
[M_4^5]&:=&-[\Kthree\times L(m;1)]+24[L(m;1,1,1)],\qquad m\text{ odd}.
\eea
Here $L(m;1,1,1)$ is the five-dimensional lens space, $\Enriques$ is the
Enriques surface, and $\Kthree$ is the K3 surface.  Again, the formulas for
$M_4^5$ are bordism identities using orientation reversal and disjoint union.
The anomaly indices in Theorem \ref{thm:main-spinc-z2m} are related to the
$\eta$-invariants of these generators by
\bea
\tilde\nu_1'(m,q)&=&\tilde a_m\eta(M_3^5,\tilde\rho_q),\cr
\tilde\nu_2'(m,q)&=&\tilde b_m\eta(M_4^5,\tilde\rho_q).
\eea
Consequently, if
$[M^5]=k_3[M_3^5]+k_4[M_4^5]$, the anomalous phase is determined by the two
integer coordinates $\tilde\nu_1'(m,q)$ and $\tilde\nu_2'(m,q)$ as in
\eqref{eq:Spin-Z2m-rewrite}.

\begin{theorem}[Integer-valued index for $\Spin\times\Z_n$]\label{thm:main-spin-zn}
Let $q\in\Z_n$ be the discrete charge of a left-handed Weyl fermion. With $a_n,b_n$ as in \eqref{eq:Spin-Zn-TP5}, the image of the continuous $\Spin\times\U(1)$ anomaly class under the homomorphism
\bea
\TP_5(\Spin\times\U(1))\longrightarrow \TP_5(\Spin\times\Z_n)
\eea
is represented by
\bea
(\nu_1'(n,q),\nu_2'(n,q))=
\left(\frac{a_n(n^2+3n+2)q^3}{6n}\bmod a_n,\;\frac{2b_n(q-q^3)}{n}\bmod b_n\right).
\eea
Equivalently, for $[M^5]=k_1[M_1^5]+k_2[M_2^5]$, the anomaly phase is the one displayed in \eqref{eq:Spin-Zn-rewrite}.
\end{theorem}

\begin{theorem}[Integer-valued index for $\Spin\times_{\Z_2^{\rF}}\Z_{2m}$]\label{thm:main-spinc-z2m}
Let $q\in\Z_{2m}$ be odd. With $\tilde a_m,\tilde b_m$ as in \eqref{eq:Spin-Z2m-TP5}, the image of the continuous $\Spin^c$ anomaly class under the homomorphism
\bea
\TP_5(\Spin^c)\longrightarrow \TP_5(\Spin\times_{\Z_2^{\rF}}\Z_{2m})
\eea
is represented by
\bea
(\tilde\nu_1'(m,q),\tilde\nu_2'(m,q))=
\left(
\tilde a_m\frac{(2m^2+m+1)q^3-(m+3)q}{48m}\bmod \tilde a_m,\;
\tilde b_m\frac{(m+1)\gcd(m+1,2)(q^3-q)}{4m}\bmod \tilde b_m
\right).
\eea
Equivalently, for $[M^5]=k_3[M_3^5]+k_4[M_4^5]$, the anomaly phase is the one displayed in \eqref{eq:Spin-Z2m-rewrite}.
\end{theorem}

The phrase ``integer-valued index'' means that the displayed coordinates lie in the cyclic factors of the finite anomaly group. The corresponding $\R/\Z$-valued $\eta$-invariants are recovered by dividing by the relevant cyclic orders.

We use the following conventions throughout.  The cyclic group $\Z_n$ is
written multiplicatively when it acts on complex vector spaces, so its elements
are roots of unity $\lambda$ with $\lambda^n=1$.  The one-dimensional
representation of charge $q$ is denoted by
$\rho_q$, meaning $\rho_q(\lambda)=\lambda^q$.  In the
$\Spin^c$-type finite symmetry
$\Spin\times_{\Z_2^{\rF}}\Z_{2m}$, the corresponding odd-charge
representation is denoted by $\tilde\rho_q$, with $q$ odd.  The symbols
$a_n,b_n,\tilde a_m,\tilde b_m$ denote the orders of the cyclic factors of
the finite anomaly groups; their prime-primary values are recalled in
\eqref{eq:Spin-Zn-TP5} and \eqref{eq:Spin-Z2m-TP5}.  The notations
$L(n;a_1,\ldots,a_k)$, $X(n;a_1,a_2)$, $\Kthree$, and $\Enriques$ are
introduced before the corresponding $\eta$-invariant computations.

\subsection{Roadmap}

In Section \ref{sec:continuous}, we review the index theorems on continuous backgrounds, deriving the classical result that captures both perturbative and mixed gravitational anomalies on $\Spin\times\U(1)$ and $\Spin^c$ manifolds.

In Section \ref{sec:eta}, we introduce the $\eta$-invariant and explain its role in the anomaly of fermions.

Section \ref{sec:Spin-U1} focuses on the discrete subgroup $\Z_n\subset\U(1)$ in the $\Spin\times\U(1)$ theory.  We compute the corresponding anomaly index by tracing the image of the cobordism group $\TP_5(\Spin\times\U(1))$ under the natural homomorphism into $\TP_5(\Spin\times\Z_n)$ and evaluating $\eta$-invariants on manifold generators.

In Section \ref{sec:Spinc}, we turn to the discrete subgroup $\Z_{2m}\subset\U(1)$ in the $\Spin^c=\Spin\times_{\Z_2^{\rF}}\U(1)$ theory. There, we derive the anomaly index for $\Spin\times_{\Z_2^{\rF}}\Z_{2m}$ by following the map from $\TP_5(\Spin^c)$ into $\TP_5(\Spin\times_{\Z_2^{\rF}}\Z_{2m})$ and evaluating $\eta$-invariants on manifold generators.

Finally, Section \ref{sec:discussion} places our results in the context of earlier work and explores their implications for discrete dark matter constructions and anomaly cancellation in discrete extensions of the Standard Model, providing a new perspective for the family puzzle (generation problem) in the Standard Model \cite{2502.21319,2512.25038,2605.26202}. We also discuss possible generalizations beyond a single cyclic symmetry.

By providing explicit formulas that descend from the continuous to the discrete setting, this paper unveils the topological underpinnings of 4d discrete anomalies and supplies a unified, computable framework for studying anomaly cancellation in theories with discrete-charged fermions.

\section{Anomaly of 4d Weyl fermion with continuous symmetry}\label{sec:continuous}

We first review the continuous anomaly indices from which the finite-symmetry
indices will be obtained.  There are two continuous symmetry types.  In the
ordinary spin case the spacetime-internal symmetry is $\Spin\times\U(1)$.  In
the $\Spin^c$ case it is
$\Spin\times_{\Z_2^{\rF}}\U(1)=\Spin^c$, where the central fermion parity in
$\Spin$ is identified with the order-two element of $\U(1)$.  By the
Atiyah--Singer index theorem \cite{Alvarez-Gaume1984,Alvarez-Gaume1985}, the
local anomaly of a four-dimensional Weyl fermion is encoded by a six-dimensional
anomaly polynomial
\bea
I_6=[\widehat A(TM)\,ch(\mathcal E)]_6 .
\eea
Here $TM$ is the tangent bundle of the six-dimensional auxiliary manifold,
$\widehat A(TM)$ is the $\widehat A$-class, $\mathcal E$ is the complex
bundle associated to the gauge representation of the fermion, and
$[\cdots]_6$ denotes the degree-six component.  We follow the standard
normalization reviewed, for example, in \cite{2302.14862,2502.21319}.

The characteristic classes used below are
\bea
\widehat A(TM)&=&1-\frac{p_1}{24}+\frac{7p_1^2-4p_2}{5760}+\cdots,\cr
ch(\mathcal E)&=&\rank(\mathcal E)+c_1(\mathcal E)
+\frac{1}{2}\bigl(c_1(\mathcal E)^2-2c_2(\mathcal E)\bigr)
\cr
&&+\frac{1}{6}\bigl(c_1(\mathcal E)^3-3c_1(\mathcal E)c_2(\mathcal E)+3c_3(\mathcal E)\bigr)+\cdots .
\eea
Here $p_i=p_i(TM)$ are Pontryagin classes of the tangent bundle and
$c_i=c_i(\mathcal E)$ are Chern classes of the gauge bundle.  For a single
left-handed Weyl fermion of $\U(1)$ charge $q$, $\mathcal E$ is a complex
line bundle.  Hence $c_i(\mathcal E)=0$ for $i>1$, and the anomaly
polynomial becomes
\bea
\exp\left(2\pi\ii \int_{M^6}I_6\right)
&=&
\exp\left(2\pi\ii\int_{M^6}[\widehat A(TM)ch(\mathcal E)]_6\right)
\cr
&=&
\exp\left(2\pi\ii\int_{M^6}\frac{q^3c_1^3}{6}-\frac{q c_1p_1}{24}\right).
\eea
In differential-form notation, if $A$ is a $\U(1)$ connection normalized by
$c_1=\dd A/(2\pi)$, then the anomaly polynomial is the exterior derivative of
a five-dimensional Chern--Simons form, $I_6=\dd I_5/(2\pi)$, with
\bea
I_5=q^3\frac{Ac_1^2}{6}-q\frac{Ap_1}{24}.
\eea
Thus, if $\partial M^6=M^5$,
\bea
\exp\left(2\pi\ii\int_{M^6}I_6\right)=\exp\left(\ii\int_{M^5}I_5\right).
\eea
The integrality of the index on closed $M^6$ implies that this five-dimensional
invertible theory is independent of the choice of filling.  Equivalently, if
$M^6$ and $M^{\prime 6}$ are two fillings of the same $M^5$, then gluing
$M^6$ to $-M^{\prime 6}$ gives a closed six-manifold, on which
$\exp(2\pi\ii\int I_6)=1$.

For a collection of Weyl fermions with charges $q$, the total invertible
field theory is obtained by summing the individual contributions:
\bea\label{eq:general}
\exp\left(\ii\int_{M^5}I_5\right)
=
\exp\left(\ii\int_{M^5}\sum_q
\left(q^3\frac{Ac_1^2}{6}-q\frac{Ap_1}{24}\right)\right).
\eea

\begin{remark}[Normalization of the continuous generators]
The continuous formulas in this section fix the normalization used throughout
the discrete computation.  In particular, the generators $I_5^A,I_5^B$ and
$\tilde I_5^C,\tilde I_5^D$ are chosen so that the coefficients of a
charge-$q$ Weyl fermion are integers: $(q^3-q)/6\in\Z$ in the
$\Spin\times\U(1)$ case and $(q^3-q)/24\in\Z$ for odd $q$ in the
$\Spin^c$ case.
\end{remark}

\subsection{\texorpdfstring{The $\Spin\times\U(1)$ case}{The Spin times U(1) case}}

In the $\Spin\times\U(1)$ case, the relevant group of five-dimensional
invertible theories is
\bea
\TP_5(\Spin\times\U(1))=\Omega_6^{\Spin\times\U(1)}\cong\Z^2.
\eea
The two integer generators can be chosen using the charge assignments
\bea
q=(1),\qquad q=(2,-1,-1),
\eea
as in \cite{2302.14862,2502.21319}.  Substituting these charge assignments into
\eqref{eq:general} gives the two generators
\bea\label{eq:I5AB}
I_5^A=\frac{Ac_1^2}{6}-\frac{Ap_1}{24},
\qquad
I_5^B=Ac_1^2.
\eea
A fermion of arbitrary integer charge $q$ then contributes
\bea
I_5=qI_5^A+\frac{q^3-q}{6}I_5^B.
\eea
The coefficient $(q^3-q)/6$ is an integer because it is the product of three
consecutive integers divided by $6$.

\subsection{\texorpdfstring{The $\Spin\times_{\Z_2^{\rF}}\U(1)=\Spin^c$ case}{The Spinc case}}

For $\Spin^c$ symmetry, fermions carry odd $\U(1)$ charge.  Geometrically,
the $\U(1)$ factor is not independent of the spin group: the element $-1\in
\U(1)$ is identified with fermion parity.  Therefore the properly normalized
integral class is the first Chern class of the quotient $\U(1)/\Z_2^{\rF}$,
which we denote by $2c_1$.  Rewriting the anomaly polynomial in terms of this
integral class gives
\bea
\tilde I_6=q^3\frac{(2c_1)^3}{48}-q\frac{(2c_1)p_1}{48}.
\eea
If $2c_1=\dd\tilde A/(2\pi)$, then $\tilde I_6=\dd\tilde I_5/(2\pi)$, with
\bea
\tilde I_5=q^3\frac{\tilde A(2c_1)^2}{48}-q\frac{\tilde A p_1}{48}.
\eea
Thus, for $\partial M^6=M^5$,
\bea
\exp\left(2\pi\ii\int_{M^6}\tilde I_6\right)
=
\exp\left(\ii\int_{M^5}\tilde I_5\right).
\eea
As above, integrality of the index on closed six-manifolds makes the
five-dimensional invertible theory well-defined.

For multiple Weyl fermions of odd charges $q$,
\bea\label{eq:general-2}
\exp\left(\ii\int_{M^5}\tilde I_5\right)
=
\exp\left(\ii\int_{M^5}\sum_q
\left(q^3\frac{\tilde A(2c_1)^2}{48}-q\frac{\tilde A p_1}{48}\right)\right).
\eea
The two generators of
$\TP_5(\Spin^c)=\Omega_6^{\Spin^c}\cong\Z^2$ may be chosen from the charge
assignments
\bea
q=(1),\qquad q=(3,-1,-1,-1),
\eea
as in \cite{2302.14862,2502.21319}.  They are
\bea\label{eq:I5CD}
\tilde I_5^C=\frac{\tilde A(2c_1)^2}{48}-\frac{\tilde A p_1}{48},
\qquad
\tilde I_5^D=\frac{1}{2}\tilde A(2c_1)^2.
\eea
A charge-$q$ fermion with $q$ odd contributes
\bea
\tilde I_5=q\tilde I_5^C+\frac{q^3-q}{24}\tilde I_5^D.
\eea
The coefficient $(q^3-q)/24$ is integral for odd $q$, because
$(q-1)q(q+1)$ is divisible by $3$, and for odd $q$ the product is also
divisible by $8$.

\section{\texorpdfstring{Introduction to the $\eta$-invariant}{Introduction to the eta-invariant}}\label{sec:eta}

We now recall the spectral invariant that computes global anomalies.  Let
$M^5$ be a closed five-manifold equipped with the relevant tangential
structure and finite background gauge field.  If a Weyl fermion transforms in a
representation $\rho$ of the finite group, we write
\bea
\eta(M^5,\rho)\in\R/\Z
\eea
for the reduced $\eta$-invariant of the five-dimensional Dirac operator
coupled to $\rho$.  With this convention, the corresponding five-dimensional
invertible field theory has partition function
\bea
\exp(-2\pi\ii\eta(M^5,\rho))\in\U(1).
\eea
Only the class of $\eta(M^5,\rho)$ modulo $\Z$ enters this phase.

The Dai--Freed description \cite{DaiFreed1994} explains why this invariant is
the correct global-anomaly phase.  If a four-dimensional fermion theory is
placed on $M^4=\partial M^5$, then its partition function is a product of the
absolute value of the four-dimensional determinant and the five-dimensional
phase,
\bea
Z_{\psi}=|\det(\cD(M^4,\rho))|\;\exp(-2\pi\ii \eta(M^5,\rho)).
\eea
Here $\cD(M^4,\rho)$ and $\cD(M^5,\rho)$ are Dirac operators coupled to the
same background data.  The Grassmann Gaussian integral gives
\bea
\int [D \bar\psi] [D\psi]\,
\exp\!\big(-\bar\psi\,\cD(M^4,\rho)\,\psi\big)
=
\det(\cD(M^4,\rho)).
\eea
The determinant has a phase ambiguity; the reduced $\eta$-invariant supplies a
canonical five-dimensional refinement of that phase.

A useful physical interpretation is obtained from a domain-wall ratio of
five-dimensional massive Dirac determinants
\cite{1508.04715,1605.02391,1607.01873,1909.08775}:
\bea
\exp(-2\pi\ii \eta(M^5,\rho))
:=\lim_{m_0\to\infty}
\frac{\det\!\big(-\ii\cD(M^5,\rho)-m_0\big)}
     {\det\!\big(-\ii\cD(M^5,\rho)+m_0\big)}
=
\lim_{m_0\to\infty}\prod_{\lambda_{M^5}}
\frac{-\ii\lambda_{M^5}-m_0}{-\ii\lambda_{M^5}+m_0},
\eea
where $\lambda_{M^5}$ runs over the eigenvalues of $\cD(M^5,\rho)$.  For a
single eigenvalue, introduce $s(\lambda_{M^5})\in(-1,1]$ by
\bea\label{eq:slambda}
\exp\!\big(-\pi\ii\,s(\lambda_{M^5})\big)
:=
\frac{-\ii\lambda_{M^5}-m_0}{-\ii\lambda_{M^5}+m_0}
=
\frac{-m_0^2+\lambda_{M^5}^2-2m_0\lambda_{M^5}\ii}
     {m_0^2+\lambda_{M^5}^2}.
\eea
As $m_0\to\infty$, $s(\lambda_{M^5})$ tends to
$\sign(\lambda_{M^5})$ for nonzero eigenvalues, while a zero eigenvalue
contributes $s(0)=1$.  Formally, this gives
\bea
\eta(M^5,\rho)
=\frac{1}{2}\left(\sum_{\lambda_{M^5}\ne0}\sign(\lambda_{M^5})
+\dim\ker\cD(M^5,\rho)\right).
\eea
The unregularized sign sum is not convergent.  The precise definition uses the
meromorphic continuation of
\bea
\eta(s):=\sum_{\lambda_{M^5}\ne0}\sign(\lambda_{M^5})|\lambda_{M^5}|^{-s},
\eea
which converges for $\Re(s)$ sufficiently large and is regular at $s=0$ in
our applications.  The reduced $\eta$-invariant is
\bea\label{eq:eta-invariant}
\eta(M^5,\rho)=\frac{1}{2}\bigl(\eta(0)+\dim\ker \cD(M^5,\rho)\bigr).
\eea
Thus $\eta(M^5,\rho)$ measures the spectral asymmetry of the five-dimensional
Dirac operator, including zero modes.  Its fractional part is the global
anomaly phase of the four-dimensional fermion.

\begin{remark}[Use of $\eta$-invariants below]
All later equalities involving $\eta$-invariants are equalities in $\R/\Z$.
By contrast, the primed quantities $\nu_i'$ and $\tilde\nu_i'$ are integer
coordinates in finite cyclic groups.  They are obtained from the
$\R/\Z$-valued $\eta$-invariants by multiplying by the order of the relevant
cyclic summand.
\end{remark}

\section{\texorpdfstring{From $\Spin\times\U(1)$ symmetry to $\Spin\times\Z_n$ symmetry}{From Spin times U(1) to Spin times Zn}}\label{sec:Spin-U1}

In this section, we restrict the continuous symmetry group $\Spin\times\U(1)$
to the finite subgroup $\Spin\times\Z_n$.  The goal is to express the anomaly
of a four-dimensional left-handed Weyl fermion of discrete charge
$q\in\Z_n$ as an element of the finite anomaly group
$\TP_5(\Spin\times\Z_n)$.  We first describe the geometric backgrounds and
spin-structure conventions, then compute the relevant $\eta$-invariants, and
finally convert the resulting $\R/\Z$-valued invariants into integer
coordinates in the two cyclic summands of $\TP_5(\Spin\times\Z_n)$.

Throughout this section, $\lambda$ denotes an element of $\Z_n$, written as an
$n$-th root of unity.  The charge-$q$ representation is the one-dimensional
character
\bea
\rho_q(\lambda)=\lambda^q,
\eea
so replacing $q$ by $q+n$ gives the same representation.  We use $RU(\Z_n)$ for
the complex representation ring of $\Z_n$:
\bea
RU(\Z_n)=\bigoplus_{q=0}^{n-1}\rho_q\Z .
\eea
If $R=\sum_q k_q\rho_q$, then $k_q$ is the number of left-handed Weyl fermions
of discrete charge $q$.

\subsection{Spin structures and geometric representatives}

We first recall the definition of a spin structure in the form needed below.
Let $E\to M$ be an oriented real vector bundle of rank $r$, equipped with a
Euclidean metric.  Denote by
\bea
P_{\mathrm{SO}}(E)\longrightarrow M
\eea
the principal $\mathrm{SO}(r)$-bundle of oriented orthonormal frames of $E$.
Let
\bea
\lambda_{\Spin}:\Spin(r)\longrightarrow \mathrm{SO}(r)
\eea
be the standard double cover.  A spin structure on $E$ is a principal
$\Spin(r)$-bundle
\bea
P_{\Spin}(E)\longrightarrow M
\eea
together with a bundle map
\bea
\theta:P_{\Spin}(E)\longrightarrow P_{\mathrm{SO}}(E)
\eea
covering the identity map on $M$ and satisfying
\bea
\theta(p\cdot s)=\theta(p)\cdot \lambda_{\Spin}(s),
\qquad
p\in P_{\Spin}(E),\quad s\in\Spin(r).
\eea
Equivalently, a spin structure is a lift of the structure group of $E$ from
$\mathrm{SO}(r)$ to $\Spin(r)$; see \cite[Ch.~II, Sec.~1]{LawsonMichelsohn1989}.
Two spin structures are equivalent if the corresponding principal
$\Spin(r)$-bundles are isomorphic over $P_{\mathrm{SO}}(E)$.  A spin structure
exists if and only if $w_2(E)=0$, and, when it exists, the equivalence classes
of spin structures form a torsor for $H^1(M;\Z_2)$; see \cite[Ch.~II, Sec.~1]{LawsonMichelsohn1989}.  For
an oriented manifold $M$, a spin structure means a spin structure on $TM$.

We now introduce the lens spaces used in the computation.  Let
\bea\label{eq:lens-space}
L(n;a_1,a_2,\dots,a_k):=S^{2k-1}/(\rho_{a_1}\oplus\rho_{a_2}\oplus\cdots\oplus\rho_{a_k})
\eea
be the $(2k-1)$-dimensional lens space.  Here $\rho_{a_i}$ is the
one-dimensional complex representation of $\Z_n$ on which $\lambda$ acts by
multiplication by $\lambda^{a_i}$.  We assume that the action is free, so each
$a_i$ is coprime to $n$.  The quotient map $S^{2k-1}\to L(n;a_1,\ldots,a_k)$
is a principal $\Z_n$-bundle, and the corresponding classifying map supplies
the natural background $\Z_n$ gauge field.  In particular,
\bea
\pi_1(L(n;a_1,\ldots,a_k))\cong \Z_n.
\eea

Let
\bea
V:=\rho_{a_1}\oplus\cdots\oplus\rho_{a_k}.
\eea
This is a complex $k$-dimensional representation of $\Z_n$.  We write $V_{\R}$
for the same vector space regarded as a real $2k$-dimensional representation,
and we identify
\bea
S^{2k-1}=S(V_{\R})
\eea
with the unit sphere in $V_{\R}$, using a $\Z_n$-invariant Hermitian metric.  The
stable tangent-bundle formula used below is first an equivariant formula over
the sphere:
\bea
TS(V_{\R})\oplus \mathbf 1_{\R}
\cong
S(V_{\R})\times V_{\R}.
\eea
Here $S(V_{\R})\times V_{\R}$ is the trivial real vector bundle over the sphere,
but it carries the induced $\Z_n$-action: $g\in\Z_n$ sends $(x,v)$ to
$(gx,gv)$.  At a point $x\in S(V_{\R})$,
\bea
T_xS(V_{\R})=\{v\in V_{\R}\mid \langle v,x\rangle=0\},
\eea
and the additional trivial line is the radial normal line spanned by $x$.  The
isomorphism is given fibrewise by
\bea
T_xS(V_{\R})\oplus\R
&\longrightarrow&
V_{\R},
\cr
(v,t)&\longmapsto& v+t x.
\eea
Since the $\Z_n$-action is unitary, it preserves the inner product and sends the
radial vector $x$ to the radial vector $gx$.  Hence this stable isomorphism is
$\Z_n$-equivariant.  Passing to the quotient gives
\bea
TL(n;a_1,\ldots,a_k)\oplus \mathbf 1_{\R}
\cong
S^{2k-1}\times_{\Z_n}V_{\R}.
\eea
The notation $S^{2k-1}\times_{\Z_n}V_{\R}$ means the real vector bundle
associated to the principal $\Z_n$-bundle $S^{2k-1}\to L(n;a_1,\ldots,a_k)$ via
the real representation $V_{\R}$.

Equivalently, if
\bea
E:=S^{2k-1}\times_{\Z_n}V
\eea
is the associated complex vector bundle, then
\bea
E_{\R}\cong S^{2k-1}\times_{\Z_n}V_{\R},
\qquad
TL(n;a_1,\ldots,a_k)\oplus\mathbf 1_{\R}\cong E_{\R}.
\eea
Adding a trivial real line does not change $w_2$, so
\bea
w_2(TL(n;a_1,\ldots,a_k))=w_2(E_{\R}).
\eea
For a complex vector bundle $E$, the underlying real bundle satisfies
\bea
w_2(E_{\R})\equiv c_1(E)\pmod 2
\eea
\cite[Sec.~14]{MilnorStasheff1974}.  Let $u\in H^2(B\Z_n;\Z)\cong\Z_n$ be the
standard generator and let $f:L(n;a_1,\ldots,a_k)\to B\Z_n$ classify the
principal $\Z_n$-bundle.  Then
\bea
c_1(E)=(a_1+\cdots+a_k)f^*u.
\eea
Writing $\bar u=f^*(u\bmod 2)$, we obtain
\bea
w_2(TL(n;a_1,\ldots,a_k))=(a_1+\cdots+a_k)\bar u.
\eea
If $n$ is odd, then $H^2(L(n;a_1,\ldots,a_k);\Z_2)=0$, so the lens space is
spin and the spin structure is unique.  If $n$ is even, freeness implies that
each $a_i$ is odd, so $a_1+\cdots+a_k\equiv k\pmod 2$.  Thus
$L(n;a_1,\ldots,a_k)$ is spin precisely when $k$ is even, except for the
one-dimensional case $k=1$, where $\bar u=0$ and $L(n;a_1)$ is spin.  In the
spin case with even $n$, there are two inequivalent spin structures because
\bea
H^1(L(n;a_1,\ldots,a_k);\Z_2)\cong\Z_2.
\eea
This agrees with the standard classification of spin structures on lens spaces;
see \cite{Franc1987}.

Five-dimensional lens spaces are not spin for all $n$, so the first universal
spin representative is instead the lens-space bundle
\bea\label{eq:lens-space-bundle}
X(n;a_1,a_2):=S(H\otimes H\oplus\mathbf{1})/\tau(a_1,a_2).
\eea
Here $H$ is the Hopf line bundle over $S^2$, $S(H\otimes H\oplus\mathbf 1)$ is
the unit sphere bundle of the rank-two complex bundle $H\otimes H\oplus\mathbf
1$, and
\bea
\tau(a_1,a_2)=\rho_{a_1}\oplus\rho_{a_2}
\eea
is the fibrewise $\Z_n$-action with weights $a_1$ and $a_2$.  When $a_1$ and
$a_2$ are coprime to $n$, the action is fixed-point free.  By \cite{Gilkey1997},
$X(n;a_1,a_2)$ carries a natural spin structure and a natural $\Z_n$-background,
with
\bea
\pi_1(X(n;a_1,a_2))\cong\Z_n.
\eea
The lift of the $\Z_n$-action to the spin bundle is determined by a square root
of the determinant character
\bea
\det(\rho_{a_1}\oplus\rho_{a_2})=\rho_{a_1+a_2}.
\eea
Equivalently, one chooses $b\in\Z/n\Z$ such that
\bea
2b\equiv a_1+a_2\pmod n.
\eea
For odd $n$, this congruence has a unique solution.  For even $n$, the numbers
$a_1$ and $a_2$ are odd, so $a_1+a_2$ is even and the congruence has exactly two
solutions.  These two solutions differ by the nontrivial class in
$H^1(X(n;a_1,a_2);\Z_2)$.  In the even case, the phrase ``positive square
root'' means that we choose one of these two lifts once and for all.

The two five-dimensional geometric backgrounds used below are
\bea
X(n;1,1)
\qquad\text{and}\qquad
L(n;1)\times \Kthree .
\eea
Here $X(n;1,1)=S(H\otimes H\oplus\mathbf 1)/\tau(1,1)$ is a lens-space bundle
over $S^2$ with fibre $L(n;1,1)$, where
\bea
\tau(1,1)(\lambda)=\diag(\lambda,\lambda).
\eea
The second background uses the one-dimensional lens space $L(n;1)=S^1/\rho_1$
and a K3 surface.  The K3 surface is spin because for any complex manifold $Y$,
\bea
w_2(TY_{\R})\equiv c_1(TY)\pmod 2,
\eea
and for $Y=\Kthree$ the canonical bundle is trivial.  Since
$c_1(K_Y)=-c_1(TY)$, we have $c_1(T\Kthree)=0$, hence
$w_2(T\Kthree)=0$.  Moreover, $\Kthree$ is simply connected, so its spin
structure is unique.  Therefore $L(n;1)\times\Kthree$ is a natural
$\Spin\times\Z_n$ background.

\subsection{\texorpdfstring{The two basic $\eta$-invariants}{The two basic eta-invariants}}

The anomaly phase of a left-handed Weyl fermion of discrete charge $q$ on a
five-dimensional $\Spin\times\Z_n$ background is
\bea
\exp(-2\pi\ii\eta(M^5,\rho_q)).
\eea
We first compute this phase on the two geometric backgrounds above.

For $X(n;1,1)$, the equivariant Lefschetz formula for the spin Dirac operator
\cite[Theorem~4.7]{Gilkey1997} gives
\bea\label{eq:X-fixed-point}
\eta(X(n;1,1),\rho_q)
=
\frac{1}{n}
\sum_{\substack{\lambda^n=1\\ \lambda\ne1}}
\lambda^q
\frac{\lambda(1+\lambda)}{(1-\lambda)^3}.
\eea
The factor $\lambda^q$ is the character of the twisting representation
$\rho_q$, while the rational function in $\lambda$ is the local fixed-point
contribution of the spin operator.

To evaluate this root-of-unity sum, compare it with the standard formula for
the seven-dimensional lens space
\bea
L(n;1,1,1,1)=S^7/\tau(1,1,1,1),
\qquad
\tau(1,1,1,1)(\lambda)=\diag(\lambda,\lambda,\lambda,\lambda).
\eea
By \cite[Theorem~4.6]{Gilkey1997},
\bea\label{eq:seven-dimensional-lens-fixed-point}
\eta(L(n;1,1,1,1),R)
=
\frac{1}{n}
\sum_{\substack{\lambda^n=1\\ \lambda\ne1}}
\Tr(R(\lambda))\frac{\lambda^2}{(1-\lambda)^4}.
\eea
For the virtual representation $R=(\rho_1-\rho_{-1})\rho_q$, one has
\bea
\Tr(R(\lambda))\frac{\lambda^2}{(1-\lambda)^4}
&=&
(\lambda^{q+1}-\lambda^{q-1})\frac{\lambda^2}{(1-\lambda)^4}
\cr
&=&
-\lambda^q\frac{\lambda(1+\lambda)}{(1-\lambda)^3}.
\eea
Therefore
\bea\label{eq:X-as-seven-dimensional-lens}
\eta(-X(n;1,1),\rho_q)
=
\eta\bigl(L(n;1,1,1,1),(\rho_1-\rho_{-1})\rho_q\bigr),
\eea
where the minus sign denotes orientation reversal.

By \cite{Gilkey1984}, the seven-dimensional lens-space invariant can be written
in terms of the $\widehat A$-polynomials:
\bea\label{eq:eta-7d-lens}
\eta(L(n;1,1,1,1),\rho_q)
=
-\frac{1}{n}
\widehat A_4\left(q+\frac{n}{2};n,1,1,1,1\right)
\mod\Z,
\eea
where
\bea
\widehat A_k(t,\vec{x})
:=
\sum_{a+2b=k}\frac{t^a\widehat A_b(\vec{x})}{a!}.
\eea
Applying this formula to $(\rho_1-\rho_{-1})\rho_q$, or equivalently taking the
corresponding finite difference, gives the cubic expression computed in
\cite{1808.02881}:
\bea\label{eq:X-eta-result}
\eta(X(n;1,1),\rho_q)
=
\frac{(n^2+3n+2)q^3}{6n}
\mod\Z.
\eea
This is the pure discrete gauge contribution.

The second invariant is the product invariant on $L(n;1)\times\Kthree$.  For an
odd-dimensional spin manifold $A$ and an even-dimensional spin manifold $B$, the
product formula gives
\bea\label{eq:eta-product}
\eta(A\times B,\rho)
=
\eta(A,\rho)\widehat A(B),
\eea
see \cite{1808.00009,Gilkey1996,Gilkey1987,Gilkey2018}.  Since
$\widehat A(\Kthree)=2$, we obtain
\bea\label{eq:eta-product-K3}
\eta(L(n;1)\times\Kthree,\rho_q)
=
2\eta(L(n;1),\rho_q).
\eea
With the spin-structure convention fixed above, \cite{1808.00009,Gilkey1996}
gives
\bea\label{eq:one-dimensional-lens-eta}
\eta(L(n;1),\rho_q)
&=&
\frac{1}{n}
\sum_{\substack{\lambda\in\Z_n\\ \lambda\ne1}}
(\lambda^q-1)\frac{\sqrt{\lambda}}{\lambda-1}
\cr
&=&
\frac{1}{n}
\sum_{\substack{\lambda\in\Z_n\\ \lambda\ne1}}
\sum_{r=1}^{q}\lambda^{\frac{2r-1}{2}}
\cr
&=&
-\frac{q}{n}
\mod\Z.
\eea
Therefore
\bea\label{eq:K3-product-eta-result}
\eta(-L(n;1)\times\Kthree,\rho_q)
=
\frac{2q}{n}
\mod\Z.
\eea
This is the linear, or mixed discrete gauge-gravitational, contribution.

We denote these two $\R/\Z$-valued invariants by
\bea\label{eq:anomaly-n}
\nu_1(n,q)
&:=&
\eta(X(n;1,1),\rho_q)
=
\frac{(n^2+3n+2)q^3}{6n}
\mod\Z,
\cr
\nu_2(n,q)
&:=&
\eta(-L(n;1)\times\Kthree,
\rho_q)
=
\frac{2q}{n}
\mod\Z.
\eea

\subsection{\texorpdfstring{Integer coordinates in $\TP_5(\Spin\times\Z_n)$}{Integer coordinates in TP5(Spin times Zn)}}

We now convert the $\R/\Z$-valued invariants into integer coordinates in the
finite anomaly group.  There is a natural restriction homomorphism
\bea
\TP_5(\Spin\times\U(1))=\Z^2
\longrightarrow
\TP_5(\Spin\times\Z_n).
\eea
It sends the two continuous generators $I_5^A$ and $I_5^B$ to pairs
$(\iota_1^A,\iota_2^A)$ and $(\iota_1^B,\iota_2^B)$, respectively.  By
\cite{1808.02881},
\bea\label{eq:Spin-Zn-TP5}
\TP_5(\Spin\times\Z_n)
=
\Omega_5^{\Spin\times\Z_n}
=
\Z_{a_n}\times\Z_{b_n}
=
\left\{\begin{array}{llll}
0&n=2,\\
\Z_n\times\Z_{\frac{n}{4}}&n=2^{\nu},\;\nu>1,\\
\Z_{3n}\times\Z_{\frac{n}{3}}&n=3^{\nu},\\
\Z_n\times\Z_n&n=p^{\nu},\;p>3\text{ is prime}.
\end{array}\right.
\eea

\begin{remark}[Prime-primary notation]
The displayed formula records the prime-primary cases.  For a general integer
$n$, one decomposes $\Z_n$ into its primary components and combines the cyclic
factors by the Chinese remainder theorem.  Equivalently, if
$n=\prod_p p^{\nu_p}$, then $a_n$ and $b_n$ are obtained by multiplying the
corresponding prime-primary orders in \eqref{eq:Spin-Zn-TP5}.  Thus the
$2$-primary part contributes $(a_{2^\nu},b_{2^\nu})=(2^\nu,2^{\nu-2})$ for
$\nu>1$, the $3$-primary part contributes $(3^{\nu+1},3^{\nu-1})$, and a prime
$p>3$ contributes $(p^\nu,p^\nu)$.  The case $n=2$ is trivial.  The formulas
below are written in terms of the resulting cyclic orders $a_n$ and $b_n$.
\end{remark}

\begin{remark}[Normalization of $\nu_2$]
The second $\R/\Z$-valued invariant is not yet the integer coordinate of the
second cyclic summand.  To obtain that coordinate, we are free to add an
integral multiple of the first invariant $\nu_1$.  Geometrically this is a
change of basis among the two five-dimensional generators; algebraically it is
the replacement of one cyclic generator by that generator plus a multiple of
the other one.  The choice made below is the one for which multiplication by
$b_n$ gives an integer-valued invariant in $\Z_{b_n}$.
\end{remark}

We seek indices $\nu_1'(n,q)\in\Z_{a_n}$ and
$\nu_2'(n,q)\in\Z_{b_n}$ such that
\bea\label{eq:condition-n}
q\iota_1^A+\frac{q^3-q}{6}\iota_1^B&=&\nu_1'(n,q),\cr
q\iota_2^A+\frac{q^3-q}{6}\iota_2^B&=&\nu_2'(n,q).
\eea
Equivalently, $(\nu_1',\nu_2')$ is the image of the continuous anomaly
\bea
qI_5^A+\frac{q^3-q}{6}I_5^B
\eea
under the restriction map.  The relations
\bea\label{eq:iotaAB}
\iota_1^A&=&\nu_1'(n,1),\cr
\iota_1^B&=&\nu_1'(n,2)+2\nu_1'(n,-1),\cr
\iota_2^A&=&\nu_2'(n,1),\cr
\iota_2^B&=&\nu_2'(n,2)+2\nu_2'(n,-1)
\eea
will be used to read off the images of $I_5^A$ and $I_5^B$.

The first coordinate is obtained by multiplying the cubic invariant by the
order of the first cyclic factor:
\bea
\nu_1'(n,q)
=
a_n\nu_1(n,q)
=
\frac{a_n(n^2+3n+2)q^3}{6n}
\mod a_n.
\eea
This is integer-valued because
\bea
\iota_1^A&=&\frac{a_n(n^2+3n+2)}{6n}\in\Z,
\cr
\iota_1^B&=&\frac{a_n(n^2+3n+2)}{n}\in\Z,
\eea
as follows from the prime-primary description of $a_n$ in
\eqref{eq:Spin-Zn-TP5}.

The second coordinate must be chosen so that it is integer-valued in
$\Z_{b_n}$.  We therefore allow a change of basis by adding an integral multiple
of $\nu_1$ to $\nu_2$:
\bea
\nu_2'(n,q)
&=&
b_n\bigl(k\nu_1(n,q)+\nu_2(n,q)\bigr)
\mod b_n
\cr
&=&
b_n\left(k\frac{(n^2+3n+2)q^3}{6n}+\frac{2q}{n}\right)
\mod b_n.
\eea
The condition that the corresponding images of $I_5^A$ and $I_5^B$ be integral is
\bea
\iota_2^A&=&b_n\left(k\frac{n^2+3n+2}{6n}+\frac{2}{n}\right)\in\Z,
\cr
\iota_2^B&=&\frac{b_nk(n^2+3n+2)}{n}\in\Z.
\eea
Using the possible values of $b_n$, the uniform choice $k=-6$ satisfies both
integrality conditions and gives the basis of integer-valued coordinates used
below.  Hence
\bea\label{eq:nu'}
(\nu_1'(n,q),\nu_2'(n,q))
=
\left(
\frac{a_n(n^2+3n+2)q^3}{6n}\mod a_n,
\frac{2b_n(q-q^3)}{n}\mod b_n
\right)
\eea
in $\TP_5(\Spin\times\Z_n)=\Z_{a_n}\times\Z_{b_n}$.  Equivalently,
\bea
I_5^A&\mapsto&
\left(\frac{a_n(n^2+3n+2)}{6n}\mod a_n,0\mod b_n\right),
\cr
I_5^B&\mapsto&
\left(\frac{2a_n}{n}\mod a_n,-\frac{12b_n}{n}\mod b_n\right).
\eea

\subsection{Geometric generators and the anomaly phase}

Set
\bea\label{eq:Spin-Zn-generators}
[M_1^5]&:=&[X(n;1,1)],\cr
[M_2^5]&:=&-[L(n;1)\times\Kthree]-6[X(n;1,1)].
\eea
Then
\bea
\nu_1'(n,q)&=&a_n\eta(M_1^5,\rho_q),\cr
\nu_2'(n,q)&=&b_n\eta(M_2^5,\rho_q).
\eea
Indeed,
\bea
\eta(M_2^5,\rho_q)
&=&
\frac{2q}{n}
-6\frac{(n^2+3n+2)q^3}{6n}
\cr
&=&
\frac{2(q-q^3)}{n}
\mod\Z,
\eea
because $(n^2+3n)q^3/n=(n+3)q^3\in\Z$.  The shift by
$-6[X(n;1,1)]$ is precisely the change of basis that removes the cubic
component from the mixed generator.

The first class detects the cubic, or pure discrete gauge, anomaly.  The second
class detects the independent mixed discrete gauge-gravitational anomaly after
this change of basis.

The fact that $M_1^5$ and $M_2^5$ generate the two cyclic summands is checked by
these normalized $\eta$-invariants.  For $q=1$,
\bea
\nu_1'(n,1)=\frac{a_n(n^2+3n+2)}{6n}
\eea
is a unit modulo $a_n$ in each nontrivial prime-primary case, and hence is a
generator of $\Z_{a_n}$; moreover $\nu_2'(n,1)=0$.  For $q=2$,
\bea
\nu_2'(n,2)=-\frac{12b_n}{n}
\eea
is a unit modulo $b_n$ in each nontrivial second factor, and hence is a generator
of $\Z_{b_n}$.  Therefore $[M_1^5]$ has exact order $a_n$ and $[M_2^5]$ has
exact order $b_n$.  Since
\bea
\Omega_5^{\Spin\times\Z_n}\cong\Z_{a_n}\oplus\Z_{b_n},
\eea
the two classes form a basis for the two cyclic summands.

Consequently, any five-dimensional $\Spin\times\Z_n$ background has a unique
coordinate expression
\bea
[M^5]=k_1[M_1^5]+k_2[M_2^5],
\qquad
k_1\in\Z_{a_n},\quad k_2\in\Z_{b_n}.
\eea
Here the group law is induced by disjoint union, and the coefficients are taken
modulo the orders of the two cyclic factors.  The coefficients $k_1$ and $k_2$
are background-dependent topological coordinates; they should not be
interpreted as numbers of fermions.  The fermion-dependent data are instead the
indices $\nu_1'(n,q)$ and $\nu_2'(n,q)$.

With these conventions, the anomaly phase on a general background is
\bea\label{eq:Spin-Zn-rewrite}
&&\exp(-2\pi\ii\eta([M^5],\rho_q))
\cr
&=&
\exp\left(
-2\pi\ii k_1\eta([M_1^5],\rho_q)
-2\pi\ii k_2\eta([M_2^5],\rho_q)
\right)
\cr
&=&
\exp\left(
-\frac{2\pi\ii}{a_n}k_1\nu_1'(n,q)
-\frac{2\pi\ii}{b_n}k_2\nu_2'(n,q)
\right)
\cr
&=&
\exp\left(
-\frac{2\pi\ii}{a_n}k_1\frac{a_n(n^2+3n+2)q^3}{6n}
-\frac{2\pi\ii}{b_n}k_2\frac{2b_n(q-q^3)}{n}
\right).
\eea

\section{\texorpdfstring{From $\Spin\times_{\Z_2^{\rF}}\U(1)=\Spin^c$ symmetry to $\Spin\times_{\Z_2^{\rF}}\Z_{2m}$ symmetry}{From Spinc to finite cyclic symmetry}}\label{sec:Spinc}

We next restrict the $\U(1)$ factor in the $\Spin^c=\Spin\times_{\Z_2^{\rF}}\U(1)$ theory to its finite subgroup $\Z_{2m}$.  The resulting finite symmetry is
$\Spin\times_{\Z_2^{\rF}}\Z_{2m}$.  Unlike the $\Spin\times\Z_n$ case, the
finite cyclic symmetry is not independent of the spin structure: the order-two
central element in $\Z_{2m}$ is identified with fermion parity.  Consequently,
fermions carry odd $\Z_{2m}$ charge.

Throughout this section, $\zeta$ denotes a primitive $2m$-th root of unity.  For
an odd integer $q$, the one-dimensional representation used to twist the Dirac
operator is
\bea
\tilde\rho_q(\zeta)=\zeta^q,
\qquad q\in 2\Z+1.
\eea
The condition that $q$ be odd means that the central element $\zeta^m=-1$ acts
as fermion parity.  We write $RU^o(\Z_{2m})$ for the additive subgroup of
$RU(\Z_{2m})$ generated by these odd-charge representations.  Thus a collection
of left-handed Weyl fermions of odd charges $q$ is represented by an element
$\sum_q k_q\tilde\rho_q\in RU^o(\Z_{2m})$.

\subsection{Finite $\Spin^c$-type structures and geometric representatives}

We first recall the tangential structure that replaces an ordinary spin
structure.  A $\Spin\times_{\Z_2^{\rF}}\Z_{2m}$-structure on an oriented
manifold $M$ can be described as a spin structure on the stabilized real bundle
\bea
TM\oplus f^*\xi.
\eea
Here $f:M\to B\Z_m$ is a classifying map, and $\xi$ is the underlying real
two-plane bundle of the complex line bundle over $B\Z_m$ induced by the
embedding $\Z_m\to\U(1)$.  The appearance of $B\Z_m$, rather than $B\Z_{2m}$,
comes from the quotient by fermion parity: after identifying the order-two
subgroup of $\Z_{2m}$ with $\Z_2^{\rF}$, the remaining determinant data is
classified by a $\Z_m$ background.  Thus $M$ admits such a structure precisely
when
\bea
w_2(TM)=w_2(f^*\xi)
\eea
for some classifying map $f$.

Let
\bea
\beta_m:H^1(-;\Z_m)\longrightarrow H^2(-;\Z)
\eea
be the Bockstein homomorphism associated with the short exact sequence
$0\to\Z\xrightarrow{\cdot m}\Z\to\Z_m\to0$.  Then
$c_1(f^*\xi)=\beta_m[f]$.  Similarly, the exact sequence
$0\to\Z_2\xrightarrow{\cdot m}\Z_{2m}\to\Z_m\to0$ gives a Bockstein homomorphism
\bea
\beta_{m,2}:H^1(-;\Z_m)\longrightarrow H^2(-;\Z_2),
\eea
and $w_2(f^*\xi)=\beta_{m,2}[f]$.  The corresponding long exact sequence in
cohomology contains
\bea\label{eq:LES}
\cdots\to H^1(-;\Z_m)\xrightarrow{\beta_{m,2}}
H^2(-;\Z_2)\xrightarrow{\cdot m}H^2(-;\Z_{2m})\to\cdots .
\eea
Consequently, if $m\,w_2(TM)=0$, then $w_2(TM)$ lies in the image of
$\beta_{m,2}$, and $M$ admits a $\Spin\times_{\Z_2^{\rF}}\Z_{2m}$-structure.

This criterion explains the geometric representatives used below.  First,
consider the Enriques surface $\Enriques$.  Its first Chern class is a nonzero
two-torsion class in $H^2(\Enriques;\Z)$, and
\bea
w_2(T\Enriques)=c_1(T\Enriques)\bmod 2
\eea
is nonzero in $H^2(\Enriques;\Z_2)$.  Therefore $\Enriques$ is not spin.  For
even $m$, however, $m\,w_2(T\Enriques)=0$, so $\Enriques$ admits a
$\Spin\times_{\Z_2^{\rF}}\Z_{2m}$-structure.  This is why the product
$L(m;1)\times\Enriques$ appears in the even-$m$ calculation.

Lens spaces also carry natural $\Spin\times_{\Z_2^{\rF}}\Z_{2m}$-structures.
For
\bea
L(m;a_1,\ldots,a_k)=S^{2k-1}/(\rho_{a_1}\oplus\cdots\oplus\rho_{a_k}),
\eea
the quotient map is a principal $\Z_m$-bundle with
\bea
\pi_1(L(m;a_1,\ldots,a_k))\cong\Z_m.
\eea
The associated complex line bundle determines the real two-plane bundle
$f^*\xi$, and the stable tangent-bundle calculation for lens spaces gives
$w_2(TL)=w_2(f^*\xi)$ in the cases used here.  Hence the lens spaces below are
natural finite $\Spin^c$-type backgrounds.

For even $m$, we use the two five-dimensional backgrounds
\bea
L(m;1,1,1)
\qquad\text{and}\qquad
L(m;1)\times\Enriques .
\eea
Here $L(m;1,1,1)=S^5/\tau(1,1,1)$, with
$\tau(1,1,1)(\lambda)=\diag(\lambda,\lambda,\lambda)$, and $L(m;1)$ is the
one-dimensional lens space.  For odd $m$, the group
$\Spin\times_{\Z_2^{\rF}}\Z_{2m}$ is equivalent to $\Spin\times\Z_m$: the
order-two subgroup of $\Z_{2m}$ splits off and is identified with fermion
parity, leaving a genuine $\Z_m$ background.  In that case we use
\bea
L(m;1,1,1)
\qquad\text{and}\qquad
L(m;1)\times\Kthree .
\eea

\subsection{The two basic $\eta$-invariants}

The anomaly phase of a left-handed Weyl fermion of odd charge $q$ on a
five-dimensional $\Spin\times_{\Z_2^{\rF}}\Z_{2m}$ background is
\bea
\exp(-2\pi\ii\eta(M^5,\tilde\rho_q)).
\eea
We now compute this phase on the two types of geometric backgrounds introduced
above.

The invariant of $L(m;1,1,1)$ is computed from the Lefschetz fixed point
formula of \cite[Theorem 4.6]{Gilkey1997} and the $\widehat A$-polynomial
formula, as in \cite{1808.02881}.  The formula is often written in terms of an
ordinary $\Z_m$-representation.  For an odd charge $q$, the corresponding
$\Z_m$-character has charge $(q-1)/2$.  Thus, if
\bea
R=\bigoplus_q k_q\rho_{\frac{q-1}{2}}\in RU(\Z_m)
\eea
corresponds to
$\tilde R=\bigoplus_q k_q\tilde\rho_q\in RU^o(\Z_{2m})$, then
\bea
\eta(L(m;1,1,1),R)
=
\frac{1}{m}
\sum_{\substack{\lambda^m=1\\ \lambda\ne1}}
\Tr(R(\lambda))\frac{\lambda^2}{(1-\lambda)^3}.
\eea
Comparing this expression with the seven-dimensional lens-space formula
\bea
\eta_{\Spin\times\Z_m}(L(m;1,1,1,1),R)
=
\frac{1}{m}
\sum_{\substack{\lambda^m=1\\ \lambda\ne1}}
\Tr(R(\lambda))\frac{\lambda^2}{(1-\lambda)^4},
\eea
one obtains
\bea
\eta(L(m;1,1,1),\tilde\rho_q)
=
\eta_{\Spin\times\Z_m}\bigl(L(m;1,1,1,1),(\rho_0-\rho_1)\rho_{\frac{q-1}{2}}\bigr).
\eea
Using \eqref{eq:eta-7d-lens}, Hsieh obtains \cite{1808.02881}
\bea\label{eq:Lm111-eta}
\eta(L(m;1,1,1),\tilde\rho_q)
=
\frac{(2m^2+m+1)q^3-(m+3)q}{48m}
\mod\Z.
\eea
This is the cubic finite $\Spin^c$-type contribution.

The second invariant is a product invariant.  The one-dimensional lens-space
calculation gives \cite{1808.00009,Gilkey1996,Gilkey2018}
\bea\label{eq:eta-L(m;1)}
\eta(L(m;1),\tilde\rho_q)
&=&
\frac{1}{2m}
\sum_{\substack{\lambda\in\Z_{2m}\\ \lambda\ne1}}
(\lambda^q-1)\frac{\lambda}{\lambda-1}
\cr
&=&
\frac{1}{2m}
\sum_{\substack{\lambda\in\Z_{2m}\\ \lambda\ne1}}
\sum_{r=1}^{q}\lambda^r
\cr
&=&
-\frac{q}{2m}
\mod\Z.
\eea
Using the product formula \eqref{eq:eta-product}, and using
$\widehat A(\Enriques)=1$, we find for even $m$ that
\bea\label{eq:E-product-eta}
\eta(-L(m;1)\times\Enriques,\tilde\rho_q)
=
\frac{q}{2m}
\mod\Z.
\eea
For odd $m$, we use $\Kthree$ instead.  Since $\widehat A(\Kthree)=2$,
\bea\label{eq:K3-product-eta-spinc}
\eta(-L(m;1)\times\Kthree,\tilde\rho_q)
=
\frac{q}{m}
\mod\Z.
\eea
It is convenient to write these two product cases in the uniform form
\bea\label{eq:anomaly-m}
\tilde\nu_1(m,q)
&:=&
\eta(L(m;1,1,1),\tilde\rho_q)
=
\frac{(2m^2+m+1)q^3-(m+3)q}{48m}
\mod\Z,
\cr
\tilde\nu_2(m,q)
&:=&
\frac{\gcd(m+1,2)q}{2m}
\mod\Z.
\eea
For even $m$, $\tilde\nu_2$ is represented by
$-L(m;1)\times\Enriques$; for odd $m$, it is represented by
$-L(m;1)\times\Kthree$.

\subsection{Integer coordinates in $\TP_5(\Spin\times_{\Z_2^{\rF}}\Z_{2m})$}

We now convert the $\R/\Z$-valued invariants into integer coordinates in the
finite anomaly group.  There is a natural restriction homomorphism
\bea
\TP_5(\Spin^c)=\Z^2
\longrightarrow
\TP_5(\Spin\times_{\Z_2^{\rF}}\Z_{2m}),
\eea
sending the continuous generators $\tilde I_5^C$ and $\tilde I_5^D$ to
$(\iota_1^C,\iota_2^C)$ and $(\iota_1^D,\iota_2^D)$, respectively.  By
\cite{1808.02881},
\bea\label{eq:Spin-Z2m-TP5}
\TP_5(\Spin\times_{\Z_2^{\rF}}\Z_{2m})
=
\Omega_5^{\Spin\times_{\Z_2^{\rF}}\Z_{2m}}
=
\Z_{\tilde a_m}\times\Z_{\tilde b_m}
=
\left\{\begin{array}{llll}
0&m=1,\\
\Z_{8m}\times\Z_{\frac{m}{2}}&m=2^{\nu}>1,\\
\Z_{3m}\times\Z_{\frac{m}{3}}&m=3^{\nu},\\
\Z_m\times\Z_m&m=p^{\nu},\;p>3\text{ is prime}.
\end{array}\right.
\eea

\begin{remark}[Prime-primary notation]
As in the $\Spin\times\Z_n$ case, the displayed group structure records the
prime-primary cases.  For a general $m$, the orders $\tilde a_m$ and
$\tilde b_m$ are obtained by multiplying the prime-primary orders.  The
$2$-primary contribution is $(8m,m/2)$ for $m=2^\nu>1$, the $3$-primary
contribution is $(3m,m/3)$, and a prime $p>3$ contributes $(m,m)$ on the
$p$-primary component.  The case $m=1$ is trivial.
\end{remark}

\begin{remark}[Normalization of $\tilde\nu_2$]
The second $\R/\Z$-valued invariant is not yet the integer coordinate of the
second cyclic summand.  To obtain that coordinate, we are free to add an
integral multiple of the first invariant $\tilde\nu_1$.  Geometrically this is a
change of basis among the two five-dimensional generators; algebraically it is
the replacement of one cyclic generator by that generator plus a multiple of
the other one.  The choice made below is the one for which multiplication by
$\tilde b_m$ gives an integer-valued invariant in $\Z_{\tilde b_m}$.
\end{remark}

We seek indices $\tilde\nu_1'(m,q)\in\Z_{\tilde a_m}$ and
$\tilde\nu_2'(m,q)\in\Z_{\tilde b_m}$ satisfying
\bea\label{eq:condition-m}
q\iota_1^C+\frac{q^3-q}{24}\iota_1^D&=&\tilde\nu_1'(m,q),\cr
q\iota_2^C+\frac{q^3-q}{24}\iota_2^D&=&\tilde\nu_2'(m,q).
\eea
Equivalently, $(\tilde\nu_1',\tilde\nu_2')$ is the image of
\bea
q\tilde I_5^C+\frac{q^3-q}{24}\tilde I_5^D
\eea
under the restriction map; see \eqref{eq:I5CD}.  The useful identities are
\bea\label{eq:iotaCD}
\iota_1^C&=&\tilde\nu_1'(m,1),\cr
\iota_1^D&=&\tilde\nu_1'(m,3)+3\tilde\nu_1'(m,-1),\cr
\iota_2^C&=&\tilde\nu_2'(m,1),\cr
\iota_2^D&=&\tilde\nu_2'(m,3)+3\tilde\nu_2'(m,-1).
\eea
The pair $(\exp(-2\pi\ii\tilde\nu_1),\exp(-2\pi\ii\tilde\nu_2))$ gives
$\R/\Z$-valued generators \cite{1808.02881}\footnote{The
$\tilde\nu_2$ in \cite{1808.02881} is our $\tilde\nu_2$ multiplied by
$\frac{m+1}{\gcd(m+1,2)}$.  Since this number is coprime to
$\tilde b_m$, the multiplication does not change the cyclic generator.}.

The first normalized invariant is obtained by multiplying the first
$\eta$-invariant by the order of the first cyclic factor:
\bea
\tilde\nu_1'(m,q)
=
\tilde a_m\tilde\nu_1(m,q)
=
\frac{\tilde a_m\bigl((2m^2+m+1)q^3-(m+3)q\bigr)}{48m}
\mod \tilde a_m.
\eea
It is integer-valued because
\bea
\iota_1^C&=&\frac{\tilde a_m(m^2-1)}{24m}\in\Z,
\cr
\iota_1^D&=&\frac{\tilde a_m(m+1)}{2m}\in\Z,
\eea
as follows from the prime-primary cases in \eqref{eq:Spin-Z2m-TP5}.

The second coordinate must be chosen so that it is integer-valued in
$\Z_{\tilde b_m}$.  We therefore allow a change of basis by adding an integral
multiple of $\tilde\nu_1$ to $\tilde\nu_2$:
\bea
\tilde\nu_2'(m,q)
&=&
\tilde b_m\bigl(k\tilde\nu_1(m,q)+\tilde\nu_2(m,q)\bigr)
\cr
&=&
\tilde b_m\left(
\frac{k\bigl((2m^2+m+1)q^3-(m+3)q\bigr)}{48m}
+
\frac{\gcd(m+1,2)q}{2m}
\right)
\mod\tilde b_m.
\eea
The integrality conditions are
\bea
\iota_2^C&=&
\tilde b_m\left(\frac{k(m^2-1)}{24m}+\frac{\gcd(m+1,2)}{2m}\right)\in\Z,
\cr
\iota_2^D&=&
\tilde b_m\frac{k(m+1)}{2m}\in\Z.
\eea
Using the possible values of $\tilde b_m$, a uniform way to satisfy the second
condition is to write $k=12l$.  The first condition is then solved by choosing
$l=\gcd(m+1,2)$; equivalently, $k=12$ for even $m$ and $k=24$ for odd $m$.
This gives the basis of integer-valued coordinates used below.  Therefore
\bea\label{eq:tildenu'}
&&(\tilde\nu_1'(m,q),\tilde\nu_2'(m,q))\cr
&=&
\left(
\frac{\tilde a_m\bigl((2m^2+m+1)q^3-(m+3)q\bigr)}{48m}\mod\tilde a_m,
\frac{\tilde b_m(m+1)\gcd(m+1,2)(q^3-q)}{4m}\mod\tilde b_m
\right)
\eea
in $\TP_5(\Spin\times_{\Z_2^{\rF}}\Z_{2m})=\Z_{\tilde a_m}\times\Z_{\tilde b_m}$.
Equivalently,
\bea
\tilde I_5^C&\mapsto&
\left(\frac{\tilde a_m(m^2-1)}{24m}\mod\tilde a_m,
0\mod\tilde b_m\right),
\cr
\tilde I_5^D&\mapsto&
\left(\frac{\tilde a_m(m+1)}{2m}\mod\tilde a_m,
\frac{6\tilde b_m(m+1)\gcd(m+1,2)}{m}\mod\tilde b_m\right).
\eea

\subsection{Geometric generators and the anomaly phase}

Define the geometric generators
\bea\label{eq:Spin-Z2m-generators}
[M_3^5]&:=&[L(m;1,1,1)],\cr
[M_4^5]&:=&-[\Enriques\times L(m;1)]+12[L(m;1,1,1)],\qquad m\text{ even},\cr
[M_4^5]&:=&-[\Kthree\times L(m;1)]+24[L(m;1,1,1)],\qquad m\text{ odd}.
\eea
Then
\bea
\tilde\nu_1'(m,q)&=&\tilde a_m\eta(M_3^5,\tilde\rho_q),\cr
\tilde\nu_2'(m,q)&=&\tilde b_m\eta(M_4^5,\tilde\rho_q).
\eea
Indeed, the coefficient $12$ for even $m$ and $24$ for odd $m$ is exactly the
change of basis described above.  It removes the part of the second
$\R/\Z$-valued invariant which is not integral after multiplication by
$\tilde b_m$.

The two classes $M_3^5$ and $M_4^5$ generate the two cyclic summands.  For
$q=1$,
\bea
\tilde\nu_1'(m,1)=\frac{\tilde a_m(m^2-1)}{24m}
\eea
is a unit modulo $\tilde a_m$ in each nontrivial prime-primary case, and hence
is a generator of $\Z_{\tilde a_m}$; moreover $\tilde\nu_2'(m,1)=0$.  For
$q=3$,
\bea
\tilde\nu_2'(m,3)
=
\frac{6\tilde b_m(m+1)\gcd(m+1,2)}{m}
\eea
is a unit modulo $\tilde b_m$ in each nontrivial second factor, and hence is a
generator of $\Z_{\tilde b_m}$.  Hence $[M_3^5]$ has exact order
$\tilde a_m$, $[M_4^5]$ has exact order $\tilde b_m$, and the two classes form
a basis for
\bea
\Omega_5^{\Spin\times_{\Z_2^{\rF}}\Z_{2m}}
\cong
\Z_{\tilde a_m}\oplus\Z_{\tilde b_m}.
\eea

Consequently, any five-dimensional $\Spin\times_{\Z_2^{\rF}}\Z_{2m}$ background
has a unique coordinate expression
\bea
[M^5]=k_3[M_3^5]+k_4[M_4^5],
\qquad
k_3\in\Z_{\tilde a_m},\quad k_4\in\Z_{\tilde b_m}.
\eea
As before, $k_3$ and $k_4$ are background-dependent topological coordinates,
not numbers of fermions.  The fermion-dependent data are the anomaly indices
$\tilde\nu_1'(m,q)$ and $\tilde\nu_2'(m,q)$.

The anomalous phase on a general background is therefore
\bea\label{eq:Spin-Z2m-rewrite}
&&\exp(-2\pi\ii\eta([M^5],\tilde\rho_q))
\cr
&=&
\exp\left(
-2\pi\ii k_3\eta([M_3^5],\tilde\rho_q)
-2\pi\ii k_4\eta([M_4^5],\tilde\rho_q)
\right)
\cr
&=&
\exp\left(
-\frac{2\pi\ii}{\tilde a_m}k_3\tilde\nu_1'(m,q)
-\frac{2\pi\ii}{\tilde b_m}k_4\tilde\nu_2'(m,q)
\right)
\cr
&=&
\exp\left(
-\frac{2\pi\ii}{\tilde a_m}k_3
\frac{\tilde a_m\bigl((2m^2+m+1)q^3-(m+3)q\bigr)}{48m}
-\frac{2\pi\ii}{\tilde b_m}k_4
\frac{\tilde b_m(m+1)\gcd(m+1,2)(q^3-q)}{4m}
\right).
\eea

\section{Discussions and applications}\label{sec:discussion}

In this section, we compare our results with the results in \cite{1808.02881,1808.00009,Gilkey1987,Gilkey1996,Gilkey2018} and discuss the applications of our results in topological quantum dark matter and the family puzzle (generation problem) in the Standard Model \cite{2502.21319,2512.25038,2605.26202} and possible generalizations beyond a single cyclic symmetry.

Our results are summarized in \Table{table:summary}.
\begin{table}[!ht]
\begin{center}
\resizebox{\textwidth}{!}{%
\begin{tabular}{|c|c|c|}
\hline
 & $\Spin\times\Z_n$ & $\Spin\times_{\Z_2^{\rF}}\Z_{2m}$\\
\hline
5d invertible theory & $\exp(-2\pi\ii\eta([M^5],\rho_q))$ & $\exp(-2\pi\ii\eta([M^5],\tilde{\rho}_q))$ \\
\hline
Manifold generators and values & $\begin{matrix}X(n;1,1) & -L(n;1)\times\Kthree\\
\exp(2\pi\ii\frac{(n^2+3n+2)q^3}{6n})&\exp(2\pi\ii\frac{2q}{n})
\end{matrix}$ & $\begin{matrix}L(m;1,1,1) & -L(m;1)\times\Enriques\text{ for even }m\\
\exp(2\pi\ii\frac{(2m^2+m+1)q^3-(m+3)q}{48m})&\exp(2\pi\ii\frac{q}{2m})\\
&-L(m;1)\times\Kthree\text{ for odd }m\\
&\exp(2\pi\ii\frac{q}{m})
\end{matrix}$\\
\hline
New manifold generators & $\begin{matrix}[M^5_1]=[X(n;1,1)]\\
[M^5_2]=-[L(n;1)\times\Kthree]-6[X(n;1,1)]\end{matrix}$ & $\begin{matrix}[M^5_3]=[L(m;1,1,1)]\\
[M^5_4]=-[L(m;1)\times\Enriques]+12[L(m;1,1,1)]\text{ for even }m\\
[M^5_4]=-[L(m;1)\times\Kthree]+24[L(m;1,1,1)]\text{ for odd }m\end{matrix}$\\
\hline
Rewrite 5d invertible theory & \eqref{eq:Spin-Zn-rewrite} & \eqref{eq:Spin-Z2m-rewrite} \\
\hline
\end{tabular}%
}
\caption{Summary of our results.}
\label{table:summary}
\end{center}
\end{table} 

For $\Spin\times\Z_n$, the generator $X(n;1,1)$ gives the cubic invariant
$\nu_1$, in agreement with the formulas of \cite{1808.02881}.  The product
$-L(n;1)\times\Kthree$ gives the linear mixed gauge--gravitational invariant,
using the product formula for $\eta$-invariants as in
\cite{1808.00009,Gilkey1987,Gilkey1996,Gilkey2018}.  The shifted generator
\bea
M_2^5=-L(n;1)\times\Kthree-6X(n;1,1)
\eea
is introduced so that the two independent cyclic coordinates in
$\Z_{a_n}\oplus\Z_{b_n}$ are separated: $M_1^5$ detects the pure
discrete gauge anomaly, while $M_2^5$ detects the independent mixed
gauge--gravitational anomaly.

For $\Spin\times_{\Z_2^{\rF}}\Z_{2m}$, the generator
$L(m;1,1,1)$ gives $\tilde\nu_1$, again matching the lens-space formulas in
\cite{1808.02881}.  The second $\R/\Z$-valued invariant is represented by
$-L(m;1)\times\Enriques$ for even $m$, and by
$-L(m;1)\times\Kthree$ for odd $m$, using
\cite{1808.00009,Gilkey1987,Gilkey1996,Gilkey2018}.  Our normalization differs
from the normalization of $\tilde\nu_2$ in \cite{1808.02881} by multiplication
by a unit in the relevant cyclic group.  This gives a minimal integer coordinate
for the second cyclic summand.

We now spell out the applications of the anomaly indices in Refs.~\cite{2502.21319,2512.25038,2605.26202}.

\subsection{Application I: symmetry extension as anomaly matching}

An anomaly that is nonzero for $G$ can sometimes be trivialized after a
symmetry extension \cite{1705.06728}.  Let
\bea
1\longrightarrow K
\longrightarrow G_{\rm Tot}
\xrightarrow{\;r\;}
G
\longrightarrow 1
\eea
be an extension of the symmetry group.  The map $r:G_{\rm Tot}\to G$ induces a
pullback map on anomaly groups,
\bea
r^*:\TP_5(G)\longrightarrow \TP_5(G_{\rm Tot}).
\eea
If an anomaly class $\alpha\in\TP_5(G)$ satisfies
\bea
r^*\alpha=0
\qquad\text{in}\qquad
\TP_5(G_{\rm Tot}),
\eea
then $\alpha$ lies in the kernel of the pullback.  The physical interpretation
used in Refs.~\cite{2502.21319,2512.25038} is that the anomalous $G$-symmetric
Weyl-fermion sector can be replaced, at low energies, by a $G$-symmetric
topologically ordered sector obtained by dynamically gauging the finite kernel
$K$.  The resulting $K$-gauge TQFT has no ordinary local particle operators
replacing the fermion field itself; instead it contains extended topological
excitations, but it carries exactly the same $G$ 't Hooft anomaly as the
original fermions.  The anomaly becomes trivial only after it is viewed as an
anomaly of the enlarged symmetry $G_{\rm Tot}$.

Ref.~\cite{2512.25038} makes this application precise in
terms of the prime-primary decomposition of the finite anomaly group.  For
charge-$q=1$ Weyl fermions, it gives the following useful
symmetry-extension rules.  For the finite $\Spin^c$-type symmetry
\bea
G=\Spin\times_{\Z_2^{\rF}}\Z_{2^{k+1}}^{\rF},
\eea
the anomaly of $2^k$ charge-one Weyl fermions is trivialized by the minimal
$K=\Z_4$ extension
\bea
1\longrightarrow \Z_4
\longrightarrow \Spin\times\Z_{2^{k+2}}
\longrightarrow \Spin\times_{\Z_2^{\rF}}\Z_{2^{k+1}}^{\rF}
\longrightarrow 1.
\eea
The smaller fermion-parity extension by $K=\Z_2^{\rF}$ does not trivialize this
$2^k$-fermion anomaly; instead, a $K=\Z_2^{\rF}$ extension trivializes the anomaly of
$2^{k+1}$ charge-one Weyl fermions through
\bea
1\longrightarrow \Z_2^{\rF}
\longrightarrow \Spin\times\Z_{2^{k+1}}
\longrightarrow \Spin\times_{\Z_2^{\rF}}\Z_{2^{k+1}}^{\rF}
\longrightarrow 1.
\eea
For the ordinary spin symmetry $G=\Spin\times\Z_{2^{k+1}}$, the anomaly of
$2^k$ charge-one Weyl fermions is trivialized by
\bea
1\longrightarrow \Z_2
\longrightarrow \Spin\times\Z_{2^{k+2}}
\longrightarrow \Spin\times\Z_{2^{k+1}}
\longrightarrow 1.
\eea
Thus the two-primary statement is not simply that one may add an arbitrary
$\Z_2$ gauge theory: the number of Weyl fermions, the fermion-parity quotient,
and the choice between $K=\Z_4$ and $K=\Z_2$ all matter.

Ref.~\cite{2512.25038} also records the odd-primary rules. For $G=\Spin\times\Z_{3^r\cdot s}$ with $2\nmid s$ and
$3\nmid s$, the $3$-primary anomaly
of charge-one Weyl fermions is trivialized for $3^r$ fermions by a minimal
$K=\Z_{3}$ extension, while an $s$-primary component is trivialized for $s$ charge-one Weyl fermions by a trivial $K=\Z_1$
extension.  The $3$-primary mechanisms apply to the group-cohomology subclass
of the finite anomaly, hence admit bosonic analogues.  In contrast, the
$2$-primary examples above are intrinsically fermionic because the role of
fermion parity in the spacetime-internal symmetry is essential.

The practical role of the present paper is to give the anomaly class
$\alpha$ in explicit integer coordinates.  For a spectrum of Weyl fermions with
charges $q_i$, one simply adds the coordinates
\bea
\left(\sum_i \nu_1'(n,q_i),\sum_i \nu_2'(n,q_i)\right)
\in \Z_{a_n}\oplus\Z_{b_n}
\eea
for $\Spin\times\Z_n$, or
\bea
\left(\sum_i \tilde\nu_1'(m,q_i),\sum_i \tilde\nu_2'(m,q_i)\right)
\in \Z_{\tilde a_m}\oplus\Z_{\tilde b_m}
\eea
for $\Spin\times_{\Z_2^{\rF}}\Z_{2m}^{\rF}$.  Vanishing of both components is
the anomaly-cancellation condition for gauging the finite symmetry.  A nonzero
pair is the anomaly that must either be matched by infrared degrees of freedom
or be trivialized by a suitable symmetry extension.  This is why the
normalization of the two cyclic coordinates is important: the proposed
$K$-gauge topological sector must match these finite-group coordinates exactly.

\subsection{Application II: topologically ordered quantum dark matter}

Refs.~\cite{2502.21319,2512.25038} apply this mechanism to Weyl fermions with the
fermionic discrete symmetries studied here,
\bea
G=\Spin\times\Z_n
\qquad\text{or}\qquad
G=\Spin\times_{\Z_2^{\rF}}\Z_{2m}^{\rF}.
\eea
The symmetry-extension problem is to find the finite kernel $K$ and the total
symmetry $G_{\rm Tot}$ for which the nonperturbative global anomaly computed
from the Atiyah--Patodi--Singer $\eta$-invariant becomes trivial after pullback.
When this happens, the anomalous Weyl fermion can be replaced, within the same
anomaly class, by a four-dimensional anomalous $G$-symmetric discrete
$K$-gauge TQFT.  The general results of Ref.~\cite{2512.25038}, summarized
above, determine which kernel is appropriate: for $G=\Spin\times_{\Z_2^{\rF}}\Z_{2m}^{\rF}$ and $m=2^p\cdot 3^r\cdot s$ 
with $2\nmid s$ and
$3\nmid s$, $K=\Z_4$ (or $K=\Z_2^{\rF}$) is minimal for the
$2$-primary branch when the number of charge-one Weyl fermions is $2^p$ (or $2^{p+1}$), $K=\Z_{3}$ is minimal for the $3$-primary branch when the number of charge-one Weyl fermions is $3^r$, and $K=\Z_1$ is trivial for the $s$-primary branch when the number of charge-one Weyl fermions is $s$.

In the Standard Model application, one considers the theory with 15 Weyl
fermions per family, omitting the 16th fermion, the sterile right-handed
neutrino $\nu_R$.  The omission of $\nu_R$ leaves a nonperturbative mixed
gauge--gravitational global anomaly involving baryon and lepton number.
Electroweak $\mathrm{SU}(2)$ instantons preserve
$\U(1)_{{\bf B}-{\bf L}}$ but break $\U(1)_{{\bf B}+{\bf L}}$ to the
fermionic discrete subgroup
$\Z^{\rF}_{2N_f,{\bf B}+{\bf L}}$.  The symmetry relevant to the family-puzzle
construction is therefore
\bea
G=
\Spin\times_{\Z_2^{\rF}}
\Z^{\rF}_{2N_f,{\bf B}+{\bf L}}.
\eea
For $N_f=3$, this becomes
\bea
G=
\Spin\times_{\Z_2^{\rF}}
\Z^{\rF}_{6,{\bf B}+{\bf L}},
\eea
and the anomaly of the three missing sterile neutrinos lies in the
$3$-primary branch of the finite anomaly classification.  Accordingly,
Ref.~\cite{2512.25038} identifies a symmetry-extension TQFT with kernel
$K=\Z_3$ that can reproduce the anomaly contribution of the three missing
$\nu_R$ fields in the infrared.  The omission of $\nu_R$ also produces a
closely related contribution to the $({\bf B}-{\bf L})$ anomaly, with the
opposite lepton-number sign; however, $\U(1)_{{\bf B}-{\bf L}}$ is not broken
to a finite subgroup by $\mathrm{SU}(2)$ instantons and is therefore not the
discrete symmetry used in the family-puzzle construction considered below.

The proposal in Refs.~\cite{2502.21319,2512.25038,2605.26202} is that the
missing anomaly can be carried instead by a symmetry-preserving $K$-gauge
fermionic TQFT, where the minimal finite gauge group is determined by the
relevant torsion branch of the finite anomaly and is not assumed at the outset
to be the color center.  This is the sense in which the topological sector is
``topologically ordered quantum dark matter'': it is a dark sector constrained
not by an ordinary perturbative gauge anomaly alone, but by the same finite
nonperturbative anomaly coordinates computed in this paper, and it does not
require an Anderson--Higgs symmetry-breaking mechanism.

\subsection{Application III: anomaly-trivializing extension and color-center matching}

Ref.~\cite{2605.26202} sharpens the preceding application by separating which
parts of a $\Z_n$ anomaly can be killed by finite symmetry extension.  The
$H^5(\Z_n,\U(1))\simeq\Z_n$ group-cohomology component, represented by the
5d cocycle
$A_{\Z_n}(\beta_{(n,n)}A_{\Z_n})(\beta_{(n,n)}A_{\Z_n})$, can be trivialized by
\bea
1\longrightarrow \Z_n\longrightarrow \Z_{n^2}\longrightarrow \Z_n\longrightarrow 1,
\eea
whereas a generic beyond-group-cohomology term involving the Pontryagin class,
written there schematically as $A_{\Z_n}p_1$, cannot be trivialized by a finite
group extension except in the special cases $n=2$ and $n=3$.

For $n=3$, one has
\bea
\TP_5(\Spin\times\Z_3)\cong\Z_9,
\qquad
a_3=9,\qquad b_3=1,
\eea
so the group-cohomology and Pontryagin-sensitive contributions are combined
into a single non-split $\Z_9$ invariant rather than appearing as two
independent cyclic coordinates.  Indeed, the coordinates derived in this paper
reduce to
\bea
\bigl(\nu_1'(3,q),\nu_2'(3,q)\bigr)
=
\bigl(q^3\mod 9,0\bigr).
\eea
Since the mod-three Pontryagin contribution
$A_{\Z_3}p_1$ vanishes, the anomaly of three charge-one Weyl fermions is the
order-three element $3\in\Z_9$.  It belongs to the group-cohomology subgroup
and is trivialized by
\bea
1\longrightarrow\Z_3
\longrightarrow\Z_9
\longrightarrow\Z_3
\longrightarrow1.
\eea

For the generalized Standard Model with $N_c$ colors and $N_f$ families,
Ref.~\cite{2605.26202} first determines the minimal cyclic gauge group
$K=\Z_N$ of the anomalous TQFT that can replace the $N_f$ missing sterile
right-handed neutrinos.  The anomaly-trivializing extension is
\bea
1\longrightarrow \Z_N
\longrightarrow \Z_{NN_f}
\longrightarrow \Z_{N_f}
\longrightarrow 1,
\eea
or, including spacetime fermion parity,
\bea
1\longrightarrow \Z_N
\longrightarrow \Spin\times\Z_{NN_f}
\longrightarrow
\Spin\times_{\Z_2^{\rF}}\Z^{\rF}_{2N_f,{\bf B}+{\bf L}}
\longrightarrow 1.
\eea
Here $N$ is determined by the anomaly and by $N_f$; it is not assumed a priori
to be the color number $N_c$.  The minimal extension order is
\bea
\left\{
\begin{array}{llll}
N=1,&N_f\ge 1,&2\nmid N_f,&3\nmid N_f,\\
N=3,&N_f\ge 3,&2\nmid N_f,&3\mid N_f,\\
N=4,&N_f\ge 2,&2\mid N_f,&3\nmid N_f,\\
N=12,&N_f\ge 6,&2\mid N_f,&3\mid N_f.
\end{array}
\right.
\eea
Thus the finite anomaly of $N_f$ charge-one fermions determines an
anomaly-trivializing $\Z_N$ gauge sector, not directly a color-center sector.
For the four-dimensional Standard Model case with $n=N_f=3$, the anomaly is
represented by a five-dimensional invertible theory, and the construction gives
a $\Spin\times\Z_3$-symmetric four-dimensional $\Z_3$-gauge TQFT that can
reproduce the anomaly contribution of the three missing sterile right-handed
neutrinos while preserving the discrete
$\Z^{\rF}_{6,{\bf B}+{\bf L}}$ symmetry.

The color-center interpretation is an additional matching condition.  If the
TQFT gauge group is to coincide with the center of the color gauge group, then
one imposes
\bea
\Z_N=\Z_{N_c}=Z(\mathrm{SU}(N_c)).
\eea
In the faithful quark/baryon notation this color-matched symmetry extension is
\bea
1\longrightarrow \Z_{N_c}
\longrightarrow
\Spin\times_{\Z_2^{\rF}}\Z^{\rF}_{2N_cN_f,{\bf Q}+N_c{\bf L}}
\longrightarrow
\Spin\times_{\Z_2^{\rF}}\Z^{\rF}_{2N_f,{\bf B}+{\bf L}}
\longrightarrow 1.
\eea
The clean Standard-Model-like identification occurs in the odd color-matched
branch, where $N_f$ and $N_c$ are odd and $N=N_c$; in this branch the baryons
are fermions and the $\Z_N$-gauge TQFT can be interpreted as the color center.
For even $N_f$, the minimal anomaly-trivializing extension contains a
$2$-primary part, $\Z_4$ or $\Z_{12}$.  One may choose an even $N_c=N$ to match
finite orders, but this is an even-color branch rather than the baryon-fermion
branch.  Therefore, only after imposing the extra odd color-matching condition
does the anomaly-trivializing extension coincide with the Standard-Model-like
color symmetry extension.  In this color-matched situation the unique minimal
nontrivial solution is
\bea
N=N_c=N_f=3.
\eea
Thus the anomaly-index formulas derived in this paper supply the arithmetic
input for the conclusion: the anomaly first selects the TQFT gauge group
$K=\Z_N$, and the familiar three-color interpretation follows only after the
additional color-center matching $N=N_c$ is imposed.

\subsection{Extensions to more general finite groups}\label{sec:generalizations}

We conclude with some comments on possible generalizations beyond a single
cyclic symmetry.  Let
\bea
A=\prod_{a=1}^r\Z_{n_a}
\eea
be a finite Abelian group, and let $T=\U(1)^r$.  The coordinatewise embedding
$\iota:A\hookrightarrow T$ induces the restriction map
\bea
\iota^*:
\TP_5(\Spin\times T)
\longrightarrow
\TP_5(\Spin\times A).
\eea
Every finite-dimensional complex representation of $A$ decomposes into
one-dimensional characters, and every such character extends to a character
of $T$.  Consequently, there is no representation-theoretic obstruction to
applying the continuous-to-discrete construction to Weyl fermions with a
finite Abelian symmetry.

More explicitly, let $q_{i,a}\in\Z$ be integer lifts of the discrete charges of
the $i$th Weyl fermion under the factors $\Z_{n_a}$.  Writing $c_1^a$ for the
first Chern class of the $a$th $\U(1)$ factor, the continuous
six-dimensional anomaly polynomial is
\bea
I_6
=
\sum_i\left[
\frac{1}{6}
\left(\sum_{a=1}^r q_{i,a}c_1^a\right)^3
-
\frac{p_1}{24}
\left(\sum_{a=1}^r q_{i,a}c_1^a\right)
\right].
\eea
The terms involving more than one index $a$ encode mixed gauge anomalies
between different cyclic factors, while the terms linear in $c_1^a$ give the
mixed gauge--gravitational anomalies.  Restricting this anomaly class along
$\iota$ therefore gives the anomaly of an arbitrary collection of Weyl
fermions transforming under $A$.

This observation concerns anomalies generated by Weyl fermions.  It does not
imply that the restriction map $\iota^*$ is surjective onto the full group
$\TP_5(\Spin\times A)$ of invertible field theories.  The bordism group
$\Omega_5^{\Spin\times A}$ can contain additional mixed torsion classes that
are not generated by free-fermion representations.  Its complete computation
may involve nontrivial differentials and extension problems in the
Atiyah--Hirzebruch or Adams spectral sequence and generally requires more
geometric representatives than the two lens-space-type generators used for a
single cyclic group.  Detecting all such classes requires a direct bordism
calculation and the evaluation of $\eta$-invariants on a sufficiently large
set of generators.

The fermionic central product can be treated similarly.  Suppose that $A$
contains a distinguished central element $z$ of order two, and choose the
embedding $\iota:A\hookrightarrow T$ so that $\iota(z)$ is the order-two
element of $T$ identified with fermion parity.  One then obtains
\bea
\iota^*:
\TP_5\left(\Spin\times_{\Z_2^{\rF}}T\right)
\longrightarrow
\TP_5\left(\Spin\times_{\Z_2^{\rF}}A\right).
\eea
The relevant fermion representations are those on which $z$ acts as $-1$.

The same general principle applies to a non-Abelian finite group $A$.  Every
finite group admits a faithful finite-dimensional unitary representation and
hence an embedding
\bea
\iota:A\hookrightarrow H
\eea
into a compact Lie group $H$, for example $H=\U(N)$.  Although the group
inclusion runs from $A$ to $H$, restriction of anomaly theories is
contravariant and gives
\bea
\iota^*:
\TP_5(\Spin\times H)
\longrightarrow
\TP_5(\Spin\times A).
\eea
If $A$ has a distinguished central involution $z$ and
$\iota(z)=z_H$ is a central involution of $H$, the corresponding map is
\bea
\iota^*:
\TP_5\left(\Spin\times_{\Z_2^{\rF}}H\right)
\longrightarrow
\TP_5\left(\Spin\times_{\Z_2^{\rF}}A\right).
\eea

Unlike the Abelian case, a representation of $A$ need not extend to a
representation of a fixed continuous group $H$.  For any specified finite set
of representations one may choose a sufficiently large continuous parent,
but there is generally no canonical choice of $H$.  Moreover, the restriction
from a chosen $H$ need not detect every element of
$\Omega_5^{\Spin\times A}$; finite non-Abelian groups can possess intrinsically
discrete anomaly classes that are not obtained from that continuous parent.
A complete classification therefore again requires a direct bordism and
$\eta$-invariant analysis.

Finally, restriction of the background symmetry requires only the inclusion
$A\hookrightarrow H$ and does not assume a physical symmetry-breaking
mechanism.  Realizing $H\to A$ dynamically additionally requires a Higgs
representation and potential whose unbroken stabilizer is $A$, which is a
model-dependent question.  These Abelian and non-Abelian generalizations
provide natural directions for future work.

\subsection*{Conclusion}

The computations above give a direct bridge from perturbative $\U(1)$ anomaly
data to finite-symmetry global anomaly data.  The final answers are expressed in
integer coordinates of the finite anomaly groups, while the derivation keeps
track of the underlying $\eta$-invariants, spin or $\Spin^c$-type structures,
and geometric manifold generators.  This makes the formulas suitable both for
comparison with bordism classifications and for practical anomaly-cancellation
tests in model building.

\section{Acknowledgments}
The author thanks Wei Cui, Pavel Putrov, Juven Wang, and Tatsuya Yamaoka for helpful comments and discussions. The author also thanks the anonymous referee for the suggestion to add \Sec{sec:generalizations}.
The author is supported by the NSFC Grant No. 12405001.
\bibliography{discrete-anomaly-PRD}

\end{document}